\documentclass[aps,prd,preprint,superscriptaddress,preprintnumbers,nofootinbib]{revtex4-1}
\pdfoutput=1

\usepackage{graphics}
\usepackage[dvips]{graphicx}
\usepackage{mathrsfs}
\usepackage{amssymb}
\usepackage{amsmath}
\usepackage{verbatim}
\usepackage{float}
\usepackage{slashed}
\usepackage{bbm}
\usepackage[dvips,letterpaper,text={6.5in,9in}]{geometry}

\usepackage[english]{babel}
\usepackage{tabu, multirow,color,dcolumn,bm,enumerate}

\newcommand{\jets}{\text{jets}}
\newcommand{\jet}{\text{jet}}
\newcommand{\tev}{\text{TeV}}
\newcommand{\gev}{\text{GeV}}
\newcommand{\fb}{\text{fb}}
\newcommand{\pb}{\text{pb}}

\usepackage{graphicx,bm}

\newcommand{\beq}{\begin{equation}}
\newcommand{\bea}{\begin{eqnarray}}
\newcommand{\eeq}{\end{equation}}
\newcommand{\eea}{\end{eqnarray}}
\newcommand{\bal}{\begin{align}}
\newcommand{\eal}{\end{align}}

\usepackage{tikz}
\usetikzlibrary{decorations.pathmorphing,decorations.markings}
\tikzset{
photon/.style={decorate, decoration={snake,amplitude=4pt, segment length=7pt}, draw=black},
particle/.style={draw=black, postaction={decorate}, decoration={markings,mark=at position .5 with {\arrow[draw=black]{>}}}},
antiparticle/.style={draw=black, postaction={decorate}, decoration={markings,mark=at position .5 with {\arrow[draw=black]{<}}}},
gluon/.style={decorate, draw=black, decoration={coil,amplitude=3pt, segment length=4pt}},
higgs/.style={draw=black,dashed,thick },
arrow/.style={draw=black, very thick, postaction={decorate}, decoration={markings,mark=at position 1 with {\arrow[draw=black]{>}}}}
}

\usepackage{latexsym}
\usepackage{mathrsfs}
\usepackage{amsthm}

\definecolor{darklightsabergreen}{rgb}{0.0, .49, 0.06}

\definecolor{NDGold}{RGB}{220, 180, 57}

\begin{document}

\title{Catching sparks from well-forged neutralinos}

\author{Joseph Bramante}
\affiliation{Department of Physics, 225 Nieuwland Science Hall, University of Notre Dame, Notre Dame, IN 46556, USA}
\author{Antonio Delgado}
\affiliation{Department of Physics, 225 Nieuwland Science Hall, University of Notre Dame, Notre Dame, IN 46556, USA}
\affiliation{Theory Division, Physics Department, CERN, CH-1211 Geneva 23, Switzerland}
\author{Fatemeh Elahi}
\affiliation{Department of Physics, 225 Nieuwland Science Hall, University of Notre Dame, Notre Dame, IN 46556, USA}
\author{Adam Martin}
\affiliation{Department of Physics, 225 Nieuwland Science Hall, University of Notre Dame, Notre Dame, IN 46556, USA}
\author{Bryan Ostdiek}
\affiliation{Department of Physics, 225 Nieuwland Science Hall, University of Notre Dame, Notre Dame, IN 46556, USA}

\preprint{CERN-PH-TH/2014-164} 

\begin{abstract}
\vspace*{0.5cm}
In this paper we present a new search technique for electroweakinos, the superpartners of electroweak gauge and Higgs bosons, based on final states with missing transverse energy, a photon, and a dilepton pair, $\ell^+\,\ell^- + \gamma + \slashed E_T$. Unlike traditional electroweakino searches, which perform best when $m_{\widetilde{\chi}^0_{2,3}} - m_{\widetilde{\chi}^0_1}, m_{\widetilde{\chi}^{\pm}} - m_{\widetilde{\chi}^0_1} > m_Z$, our search favors nearly degenerate spectra; degenerate electroweakinos typically have a larger branching ratio to photons, and the cut $m_{\ell\ell} \ll m_Z$ effectively removes on-shell Z boson backgrounds while retaining the signal. This feature makes our technique optimal for `well-tempered' scenarios, where the dark matter relic abundance is achieved with inter-electroweakino splittings of $\sim 20 - 70\,\gev$. Additionally, our strategy applies to a wider range of scenarios where the lightest neutralinos are almost degenerate, but only make up a subdominant component of the dark matter -- a spectrum we dub `well-forged'. Focusing on bino-Higgsino admixtures, we present optimal cuts and expected efficiencies for several benchmark scenarios. We find bino-Higgsino mixtures with $m_{\widetilde{\chi}^0_{2,3}} \lesssim 190\,\gev$ and $m_{\widetilde{\chi}^0_{2,3}} - m_{\widetilde{\chi}^0_1} \cong 30\,\gev$ can be uncovered after roughly $600\,\fb^{-1}$ of luminosity at the 14 TeV LHC\@. Scenarios with lighter states require less data for discovery, while scenarios with heavier states or larger mass splittings are harder to discriminate from the background and require more data. Unlike many searches for supersymmetry, electroweakino searches are one area where the high luminosity of the next LHC run, rather than the increased energy, is crucial for discovery.
\end{abstract}

\maketitle

\section{Introduction}
\label{sec:intro}

Supersymmetry has been a scenario strenuously sought at the LHC. While the lack of signals from standard searches has put several constraints on the spectrum of superpartners, and especially on the colored superpartners, these standard searches and channels have problems dealing with compressed supersymmetric mass spectra \cite{Strassler:2006qa,Fan:2011yu,Fan:2012jf,Aad:2014nra}. One circumstance in which several superpartners with similar mass are expected is neutralino dark matter (DM). It is well known that the MSSM with R-parity has candidates that explain the relic density of DM, the lightest neutralino being perhaps the most natural candidate. Taking into consideration all available experimental data, one finds that for pure states,  bino, wino or Higgsino, there are  difficulties accommodating the measured dark matter relic density because, either the bino does not interact sufficiently and overcloses the universe or Higgsinos and winos annihilate too efficiently and have to be over a TeV and thus outside LHC detection range, to explain the DM density. On the other hand, a non-trivial mixture (i.e. well-tempered) of the bino and the Higgsino, or the bino and wino, or all three, can reproduce the measured DM abundance with masses in the hundreds of GeV \cite{Feng:2000gh,BirkedalHansen:2001is,ArkaniHamed:2006mb,Cheung:2012qy,Cohen:2013kna,Han:2013gba,Henrot-Versille:2013yma,Huang:2014xua,Katz:2014mba}. 

Naturalness is also a powerful guide to the mass spectrum of beyond the standard model (BSM) scenarios \cite{Barbieri:1987fn,Kitano:2006gv,Papucci:2011wy,Brust:2011tb,Baer:2012up,Baer:2012cf,Baer:2013gva,Kribs:2013lua,Farina:2013ssa,Boehm:2013qva,Baer:2014cua}. As the Higgsino mass parameter $\mu$ enters at tree level into the expression for the Higgs mass, Higgsinos must be near the weak scale, $O(200\,\gev)$, to remain natural\footnote{Unless the relationship between $\mu$ and other Higgs soft mass parameters is fixed by some UV dynamics~\cite{Roy:2007nz}.}. Other superpartner masses, such as the bino mass $M_1$ and the wino mass $M_2$ also contribute to the Higgs mass, though at loop level. Therefore the bino and wino may be significantly heavier than the weak scale while remaining natural.

Viewed in the light of naturalness, the bino-Higgsino admixture stands out among other well-tempered scenarios and is a prime target for LHC searches. This admixture will be the focus of this paper, with the study of well-tempered possibilities involving the wino deferred to later work. Well tempered bino-Higgsino scenarios are characterized by small inter-electroweakino splittings; in terms of Lagrangian parameters, well tempered bino-Higgsinos with $|\mu| < 200 ~{\rm GeV}$ have
\begin{equation}
M_1 \simeq |\mu| - 25 ~{\rm GeV},
\end{equation}
where $M_1$ is the bino soft mass parameter and $\mu$ is the Higgsino mass. Translated into mass eigenvalues, the above relation implies the splitting between the lightest neutralino $\widetilde{\chi}^0_1$ (the lightest supersymmetric particle (LSP)) and the next two neutralinos $\widetilde{\chi}^0_2, \widetilde{\chi}^0_3$, as well as the splitting between the lightest chargino $\widetilde{\chi}^{\pm}_1$ and the LSP are all $\lesssim m_Z$. 

The combination of the light bino-Higgsino neutralino sector masses, preferred by naturalness arguments, and the small inter-state splitting puts the electroweakino sector of these models in a confounding place; the states are light enough to be produced abundantly at the LHC, but the small splitting among states makes conventional analyses difficult. Conventional analyses, assuming all sleptons are heavier than the electroweakinos, are based on the trilepton plus missing energy signal, $pp \to 3\ell + \slashed E_T$. This final state is generated by the production of heavier electroweakinos $pp \to \widetilde{\chi}^{\pm}_1\,\widetilde{\chi}^0_2$, followed by the decays $\widetilde{\chi}^{\pm}_1 \to W^{\pm}(\ell^{\pm}\nu) + \widetilde{\chi}^0_1$, $\widetilde{\chi}^0_2 \to Z(\ell^+\ell^-) + \widetilde{\chi}^0_1$. As $m_{\widetilde{\chi}^0_{2,3}} - m_{\widetilde{\chi}^0_1}$ and $m_{\widetilde{\chi}^{\pm}_1} - m_{\widetilde{\chi}^0_1}$ fall below $m_Z$, the sensitivity of this approach degrades; the intermediate $W^{\pm}$ and $Z$ bosons become off-shell and their subsequent lepton decays are too soft to trigger upon efficiently.

One way to combat the loss of sensitivity in the trilepton plus $\slashed E_T$ channel is to look for electroweakinos produced in association with a hard photon or jet, $pp \to \widetilde{\chi}\widetilde{\chi} + j/\gamma$. Since the initial state radiation (ISR) can be used as a triggerable object, rather than the electroweakino decay products, the subsequent cuts can be loosened, opening up sensitivity to smaller electroweakino mass splittings. The price one pays for this approach is a significant loss in rate. The decrease depends on the jet and photon trigger thresholds, but is roughly $1/50$ for electroweakinos produced with a $100\,\gev$ jet at a 14 TeV LHC \cite{Schwaller:2013baa,Baer:2014cua,Han:2014kaa}. 

In this paper we present an alternative analysis strategy for nearly degenerate electroweakinos that does not rely on large missing tranverse momentum concomitant with hard initial state radiation. Instead, we look to a different final state, $\ell^+ \ell^- + \gamma + \slashed E_T$. Electroweakino final states containing photons can certainly arise from initial or final-state radiation, such as $p p \to \widetilde{\chi}^+ \widetilde{\chi}^- + \gamma \to \ell^+ \ell^- + \gamma + \slashed E_T$, however photons can also come from neutralino decay \cite{Ambrosanio:1996gz,Baer:2002kv,Baer:2005ky}, $\widetilde{\chi}^0_{2,3} \to \widetilde{\chi}^0_1 + \gamma$. Because decays to photons are a loop-level effect and are therefore sometimes neglected in electroweakino phenomenology. Indeed, Tevatron studies have placed bounds on neutralino masses in gauge-mediated scenarios by searching for their decays to photons, Z bosons, and gravitinos \cite{Abazov:2004jx,Aaltonen:2009tp,Abazov:2012qka}. Photon decays are two body processes and can easily compete with three-body decays through an off-shell electroweak gauge boson, such as $\widetilde{\chi}^0_{2,3} \to \ell^+ \ell^- \widetilde{\chi}^0_1$. As we will show, photons from neutralino decays can yield a more easily distinguished signal in future high luminosity collider data.

Assuming $\widetilde{\chi}^0_{2,3} \to \gamma + \widetilde{\chi}^0_1$ makes up a portion of our electroweakino signal, the next question is what else is present. One possibility is that $\widetilde{\chi}^0_{2,3}$ is produced in association with a chargino, $pp \to \widetilde{\chi}^{\pm}\widetilde{\chi}^0_{2,3}$, which leads to a final state of $\ell + \gamma + \slashed E_T$\footnote{Throughout this work we will focus on electroweakino decays that yield leptons. The backgrounds for hadronic final states are orders of magnitude larger, especially considering the low energies (i.e small splittings) we are interested in.}. This process has the benefits of a large production rate  relative to other electroweakino processes and the $O(100\%)$ branching fraction of the chargino to $W^*$. However, this final state has a large SM background from $W(\ell \nu) + \gamma$, produced via $\sigma(p p \to W^{\pm}(\ell\nu)\gamma) \sim 30\,\pb$ at the LHC  ($\sqrt{s} = 14\,\tev$). While there are certainly kinematic handles that can distinguish $W + \gamma$ from $\widetilde{\chi}^{\pm} \widetilde{\chi}^0_{2,3}$ production, the starting diboson cross section is so enormous ($> 100$ times the signal) that an electroweakino search using only a single-lepton final states looks extremely challenging.  

We therefore focus on the final state $\ell^+ \ell^- + \gamma + \slashed E_T$. While there are many electroweakino production and decay paths that arrive at this final state, we find $pp \to \widetilde{\chi}^0_3 \widetilde{\chi}^0_2,\, \widetilde{\chi}^0_{2,3} \to \gamma \slashed E_T,\, \widetilde{\chi}^0_{3,2} \to \ell^+ \ell^- \slashed E_T$ -- one neutralino decays to a photon plus LSP and the other decays to a same-flavor lepton pair via an off-shell $Z$ -- has the best combination of rate and kinematic discernibility from SM processes. One immediate benefit of the $\ell^+\ell + \gamma + \slashed E_T$ final state is that there is no diboson background. There are formidable backgrounds coming from $pp \to VV\gamma$, where $V$ are any combination of $W^{\pm}/Z/\gamma^*$, and from $pp \to \gamma^*/Z(\tau^+\tau^-) + \gamma$ where both of the taus decay leptonically. However, we find the signal can be separated from the SM using a combination of $m_{\ell\ell}$ and angular cuts.

The layout of the remainder of this paper is as follows: In Sec. \ref{sec:fullspec} we review the existing limits on Higgsino-bino admixtures, then explore how the inter-electroweakino splitting, the overall electroweakino mass scale, and the relic density are interrelated. Next, In Sec. \ref{sec:subSignal} we introduce the $\ell^+\ell^- \gamma + \slashed E_T$ final state and study its rate and kinematic properties. Our main results are presented in Sec.~\ref{sec:collid} where we motivate and implement an analysis that isolates the Higgsino-bino signal from the SM background. We first test our analysis on four benchmark points, then, discuss how our strategy fares in a wider region of parameter space. In Sec.~\ref{sec:discussion} we comment on how our fairly idealized setup holds up under more realistic experimental conditions. Finally, Sec.~\ref{sec:conc} contains our conclusions. Some technical details can be found in the appendices.

\section{The Mass Splitting and Relic Abundance of bino-Higgsinos}
\label{sec:fullspec}

In this section we determine bino-Higgsino mass splittings and relic abundances for the mass parameter ranges $M_1 = 100-250 ~{\rm GeV}$, $|\mu| = 100-250 ~{\rm GeV}$, and for $\tan\beta = 2~{\rm and}~10$. Here and throughout this paper, it is assumed that the wino (mass parameter $M_2$), and all other supersymmetric particles are decoupled; their masses are set to $\sim 3 ~{\rm TeV}$ in numerical computations.  We do not explicitly determine how a large Higgs mass is generated, but the model building details which yield $m_h = 125 ~{\rm GeV}$ should not affect the results of this paper\footnote{While we do not specify how the mass of the Higgs is generated, heavy stops and F or D term contributions could be responsible for raising the Higgs mass. In any case, in this study the stop and gluino are assumed to completely decouple from electroweakinos and we ignore them in our treatment of the electroweakino mass spectrum.}. The treatment of well-tempered neutralinos given below -- masses, collider and dark matter properties -- can be easily adapted to mixed bino-wino scenarios, which we leave to future work. 

\subsection{Status of bino and Higgsino collider searches}
\label{sec:status}

Many of the most stringent bounds on the bino, Higgsino, and bino-Higgsino admixture were set over a decade ago by LEP and LEPII. The exclusion of charginos produced via $e^+ e^- \rightarrow \widetilde{\chi}^+_1 \widetilde{\chi}^-_1$ at LEPII bounds the lightest chargino mass, $m_{\widetilde{\chi}^{\pm}} > 103 ~{\rm GeV}$ \cite{lepii}; since we have decoupled the wino, in our setup this limit is essentially a limit on $|\mu|$. Recent multilepton plus $\slashed E_T$ studies at the LHC have restricted some mixed neutralino parameter space \cite{Aad:2014nua,Chatrchyan:2014aea}. As we have decoupled all sleptons, the limits that apply are for electroweakinos that decay via $W^{(*)\pm}/Z^{(*)}$ and generate a $3\ell + \slashed E_T$ final state. When $m_{\widetilde{\chi}^0_2} - m_{\widetilde{\chi}^0_1}$ is greater than $m_Z$, $\widetilde{\chi}^0_2$ (assumed degenerate with $\widetilde{\chi}^{\pm}_1$) are excluded up to $400\,\gev$ for massless $\widetilde{\chi}^0_1$ and up to $350\,\gev$ for $\widetilde{\chi}^0_1$ lighter than $\sim 150\,\gev$. For more degenerate spectra, $m_{\widetilde{\chi}^0_2} - m_{\widetilde{\chi}^0_1}, m_{\widetilde{\chi}^{\pm}_1} - m_{\widetilde{\chi}^0_1} < m_Z$, the bounds are even weaker, with no limits for $m_{\widetilde{\chi}^0_1} > 100\,\gev$. 
	
Looking forward to the 14 TeV LHC run, the sensitivity of $3\ell + \slashed E_T$ searches to scenarios with $m_{\widetilde{\chi}^0_2} - m_{\widetilde{\chi}^0_1}, m_{\widetilde{\chi}^{\pm}_1} - m_{\widetilde{\chi}^0_1} > m_Z$ will extend greatly~\cite{CMS:2013xfa,ATL-PHYS-PUB-2013-002}, but the sensitivity to nearly degenerate spectra will not. New collider techniques to search this area of neutralino parameter space are necessary. This region where the electroweakino spectra is quasi-degenerate also has a compelling connection with dark matter -- as we will see, well-tempered bino-Higgsino scenarios with minimal fine-tuning ($|\mu| \lesssim 200~\gev$) typically have inter-electroweakino mass splittings of $O(25~\gev)$.

\subsection{Bino-Higgsino mass splitting}
\label{sec:splitting}
  
Given the insensitivity of current LHC  searches to sub-$m_Z$ inter-electroweakino splitting, our next step is to analyze what regions of bino-Higgsino parameter space lead to quasi-degenerate spectra. Under our assumption of a decoupled wino, the mass eigenstates of the bino-Higgsino depend on the  masses $M_1,\, \mu,\, m_Z$ and the angles $\theta_W$ and $\beta$.  In a basis with a column vector, which from top to bottom has first the bino $\tilde{B}$ and then the Higgsino mass states defined as $\tilde{H_1} \equiv (\tilde{H}_u-\tilde{H}_d)/\sqrt{2}$ and $\tilde{H_2} \equiv (\tilde{H}_u+\tilde{H}_d)/\sqrt{2}$, the mass mixing matrix is given by
\begin{align}
\mathcal{M} = \left(\begin{smallmatrix} M_1 & -\frac{s_\beta +c_\beta}{\sqrt{2}}s_W m_Z&
 \frac{s_\beta -c_\beta}{\sqrt{2}}s_W m_Z \\
-\frac{s_\beta +c_\beta}{\sqrt{2}}s_W m_Z &\mu &0 \\
\frac{s_\beta -c_\beta}{\sqrt{2}}s_W m_Z & 0 & -\mu \end{smallmatrix} \right),
\label{eq:mmatrix}
\end{align}
where $s_\beta$ and $c_\beta$ represent $\sin\beta$ and $\cos\beta$ respectively, and $s_W$ is $\sin\theta_W$.

\begin{figure}[h!]
\begin{center}
\includegraphics[scale=1]{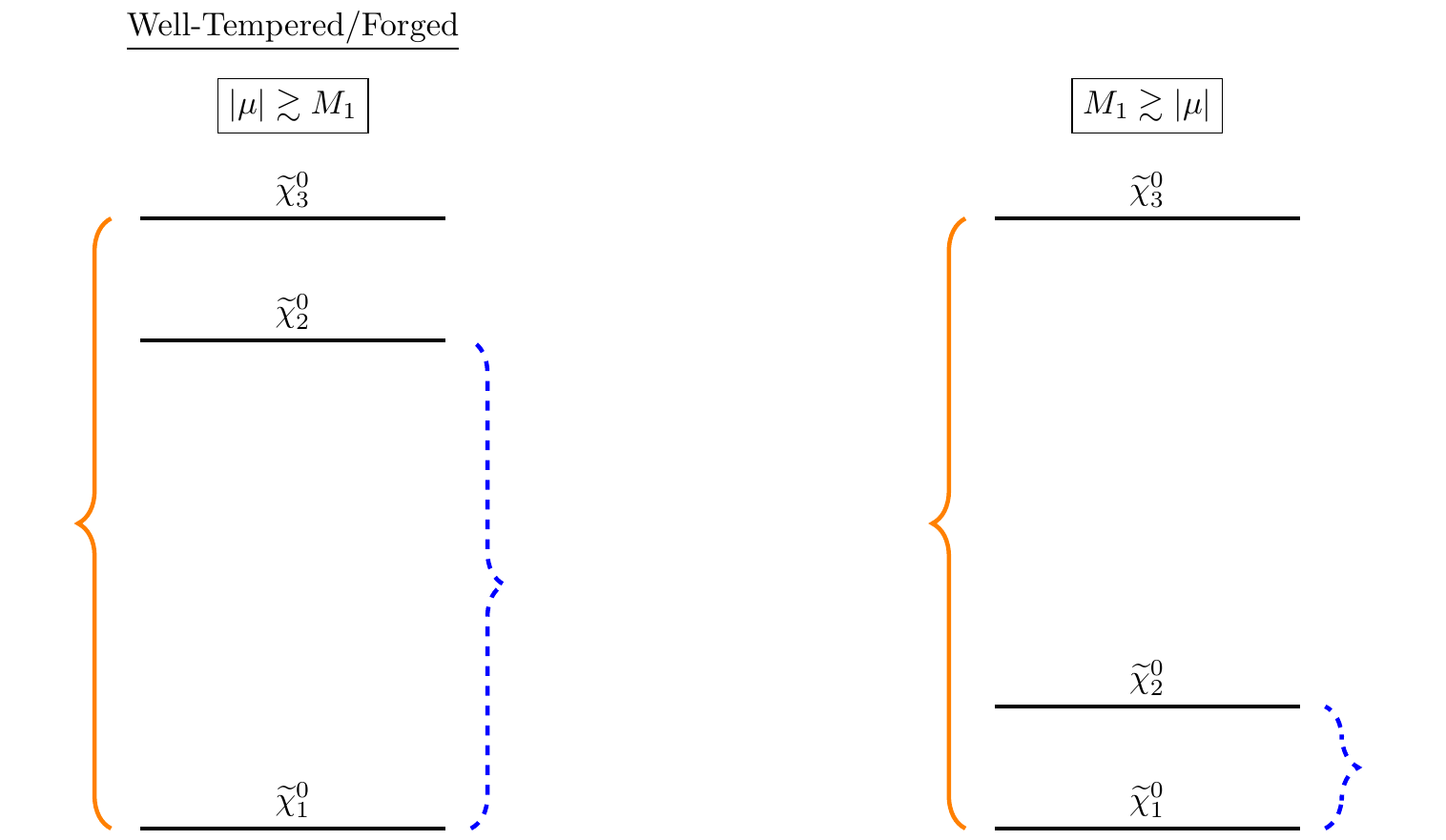}
\caption{The mass splitting for different ranges of $|\mu|$ and $M_1$. On the left side, $M_1 < |\mu|$ and so the two Higgsino-like states are heavier than the bino-like state. This corresponds to the lower, outside edge of the plots shown in Fig.~\ref{fig:params1}. Both the splitting $m_{\widetilde{\chi}^0_3}-m_{\widetilde{\chi}^0_1}$ and $m_{\widetilde{\chi}^0_2}-m_{\widetilde{\chi}^0_1}$ are large, making this spectrum amenable to studies with a photon and dilepton pair in the final state. The right side shows the opposite regime, where $|\mu| < M_1$. This results in the Higgsinos having smaller masses than the bino, and is shown in the upper, inside edge of the plots in Fig.~\ref{fig:params1}. In this case the splitting $m_{\widetilde{\chi}^0_3}-m_{\widetilde{\chi}^0_1}$ is large and the splitting $m_{\widetilde{\chi}^0_2}-m_{\widetilde{\chi}^0_1}$ is small.}
\label{fig:masssplit}
\end{center}
\end{figure}

The masses and composition of the neutralinos determine how they are produced and decay. To see how the masses and mass splittings $m_{\widetilde{\chi}_{2,3}^0} - m_{\widetilde{\chi}_1^0}$ shift as we vary the relationship between $|\mu|$, $M_1$ and $\tan{\beta}$, it is useful to first explore some limiting cases. If $|\mu| \gg M_1$, the heaviest two neutralinos $\widetilde{\chi}_2^0$ and $\widetilde{\chi}_3^0$ will be Higgsino like, and the mass splittings between each of these heavy neutralinos and the bino-like LSP $\widetilde{\chi}_1^0$ will be sizable. This spectrum is shown in the left-hand panel of Fig.~\ref{fig:masssplit}.  If, on the other hand, $|\mu| \ll M_1$ (sample spectrum shown in the right panel of Figure \ref{fig:masssplit}), the lightest two neutralinos $\widetilde{\chi}_1^0$ and $\widetilde{\chi}_2^0$ will be Higgsino like. This means that while the mass splitting between $\widetilde{\chi}_3^0$ and $\widetilde{\chi}_1^0$ will remain sizable as before, the mass splitting between $\widetilde{\chi}_2^0$ and $\widetilde{\chi}_1^0$ will be tiny. 

To interpolate between these limits, we proceed numerically. The mass splittings for $|\mu |, M_1 \in [100\,\gev, 250\,\gev ]$ are shown below in Fig.~\ref{fig:params1}. In the upper right (left) of the $\mu < 0 $ ($\mu > 0)$ plots, we see the limit $M_1 \gg |\mu|$, while we see the $|\mu| \gg M_1$ limit in the lower left (right) corner. In the middle region, where $|\mu| \sim M_1$, the neutralino spectrum becomes compressed and there are broad regions of parameter space where either $m_{\widetilde{\chi}^0_2}-m_{\widetilde{\chi}^0_1}$,  $m_{\widetilde{\chi}^0_3}-m_{\widetilde{\chi}^0_1}$, or both are less than the $Z$ mass; these slices are the regions of greatest interest for our study. While the splittings are less than $m_Z$ in these slices, they do not become arbitrarily small. In the region $|\mu| \sim M_1$, the splitting between $\widetilde{\chi}^0_3$ and $\widetilde{\chi}^0_1$ is always greater than $\sim 30\,\gev$. The $\widetilde{\chi}^0_2 - \widetilde{\chi}^0_1$ splitting can be smaller, i.e. in the $|\mu| \ll M_1$ limit, but for $M_1 \sim |\mu|$ the splitting rarely dips below $\sim 20\,\gev$. As we explain in greater detail in Sec.~\ref{sec:collid}, splittings of this size are interesting from a collider perspective; the splittings are large enough that particles emitted as one neutralino decays to another can be efficiently detected at the LHC, yet the splittings are too small for neutralinos to decay via on-shell $W^{\pm}/Z$.

 Comparing the four plots in Fig.~\ref{fig:params1}, we see that the degree of degeneracy depends on the sign of $\mu$ and $\tan{\beta}$. The off-diagonal matrix entries in Eq.~(\ref{eq:mmatrix}) grow in magnitude with $\tan{\beta}$, thus the inter-state splitting also increases with $\tan{\beta}$. To understand the effect of the sign of $\mu$, consider the limit that $M_1\sim  \mu$, and $\tan \beta =1$. In this case, $ m_{\widetilde{\chi}_3^0} - m_{\widetilde{\chi}_2^0} \sim m_{\widetilde{\chi}_2^0} - m_{\widetilde{\chi}_1^0} = \frac{1}{2}\left( m_{\widetilde{\chi}_3^0} - m_{\widetilde{\chi}_1^0} \right) \sim m_W \tan \theta_W.$ However, for $M_1 \sim - \mu$, we find $ m_{\widetilde{\chi}_3^0} \sim m_{\widetilde{\chi}_2^0} > m_{\widetilde{\chi}_1^0}$. The splitting between $m_{\widetilde{\chi}_2^0} - m_{\widetilde{\chi}_1^0}$ for $\mu$ positive is greater than $\mu$ negative. Hence, as reflected in the left and right halves of Fig.~\ref{fig:params1}, the mass splittings $m_{\widetilde{\chi}_{2,3} ^0} -m_{\widetilde{\chi}_1^0}$ for a positive $\mu$ are greater than that of a negative $\mu$. Combining these trends, the smallest inter-neutralino splittings occur when $\tan{\beta}$ is small and $\mu < 0$ while the splittings are largest for large $\tan{\beta}, \mu > 0$. 
 
 Finally, we note that the well-tempered/forged bino-Higgsino chargino mass, when $|\mu| > M_1$, will be very nearly the mass of $\widetilde{\chi}^0_2$.

\subsection{Bino-Higgsino relic abundance}

The inter-neutralino mass splittings also have ramifications for neutralino dark matter relic abundance, since the lightest neutralino is assumed to be stable. Before describing how the $M_1 - \mu$ split affects mixed bino-Higgsino relic abundance, we note that a lightest neutralino that is purely bino or Higgsino does not make a viable electroweak-scale ($100\,\gev - 1\,\tev$) dark matter candidate. In order for a pure bino LSP to be viable it must coannihilate with another sparticle\footnote{Coannihilating bino DM is limited by LEP sfermion constraints, e.g. $m_{\tilde{\ell^\pm}} > {\rm 100 ~GeV} $. Assuming bino coannihilation with a right handed slepton, in the limit $m_{\tilde{B}} < m_{\tilde{\ell^\pm}}$, the bino relic,
$
 \Omega_{\tilde{B}} h^2 \simeq 10^{-2} \left( m_{\tilde{\ell^\pm}}^2 / (\rm 100 ~ GeV)  \right)^2/M_1^2,$
indicates that the bino must be lighter than $\sim {\rm 50 ~GeV}$, a mass range constrained by the Z width.}, while a purely Higgsino LSP can be viable only if its mass is fine-tuned to be greater than $1\,\tev$, Refs.~\cite{ArkaniHamed:2006mb,Perelstein:2011tg,Perelstein:2012qg,Cheung:2012qy,Huang:2014xua}.

\begin{figure}[h!]
\includegraphics[width=0.45\textwidth]{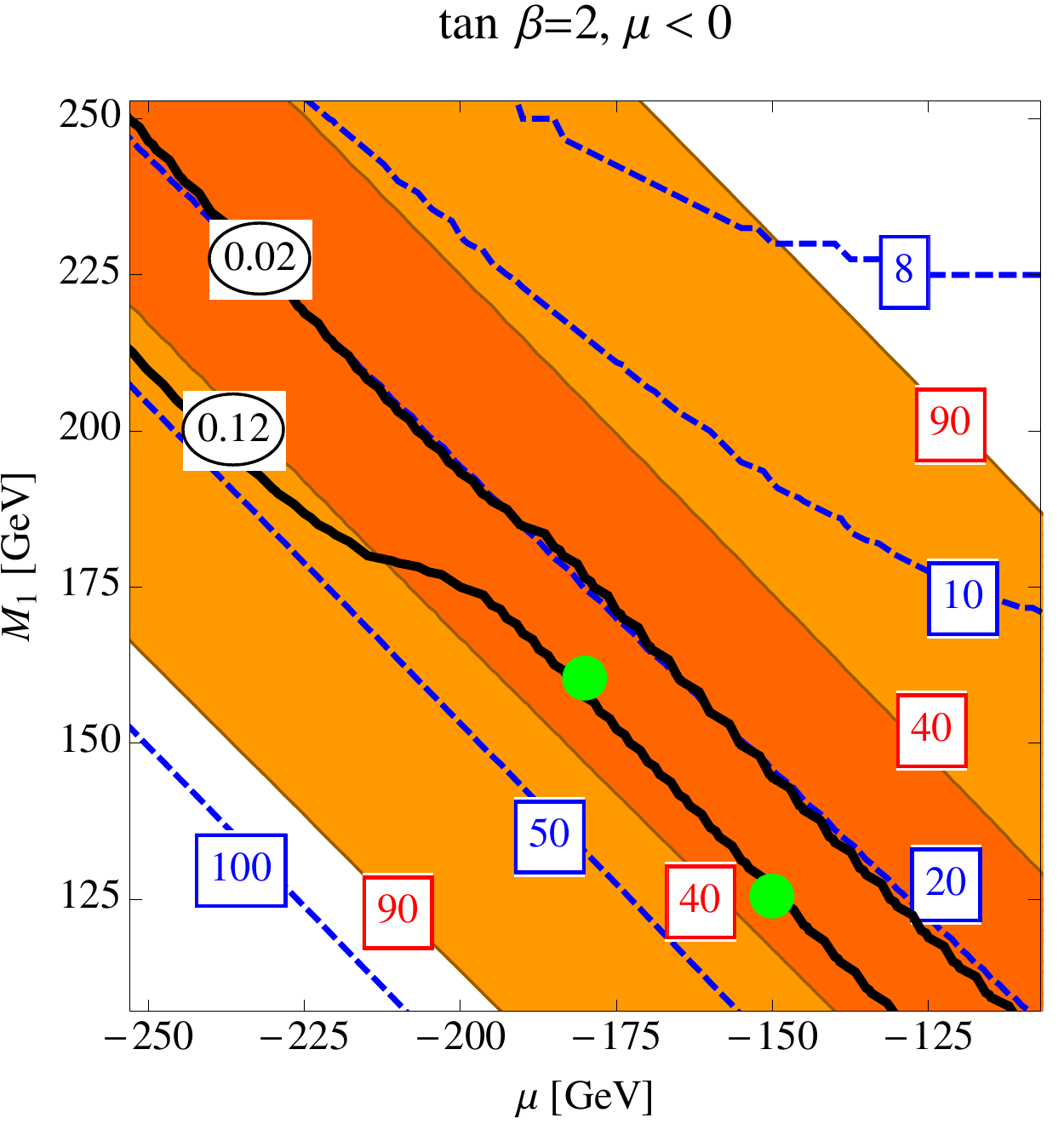}
\includegraphics[width=0.45\textwidth]{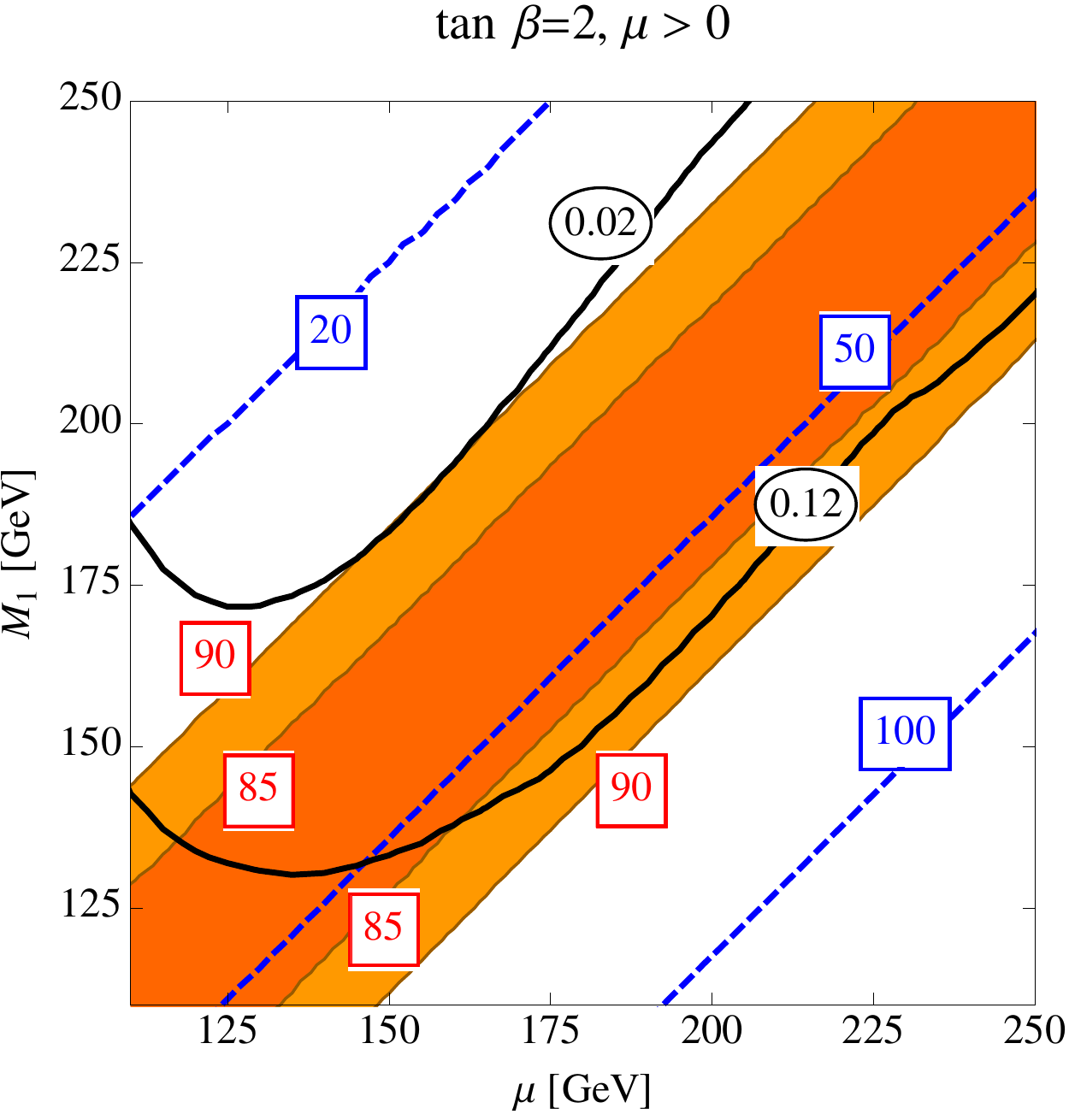}\\
\includegraphics[width=0.45\textwidth]{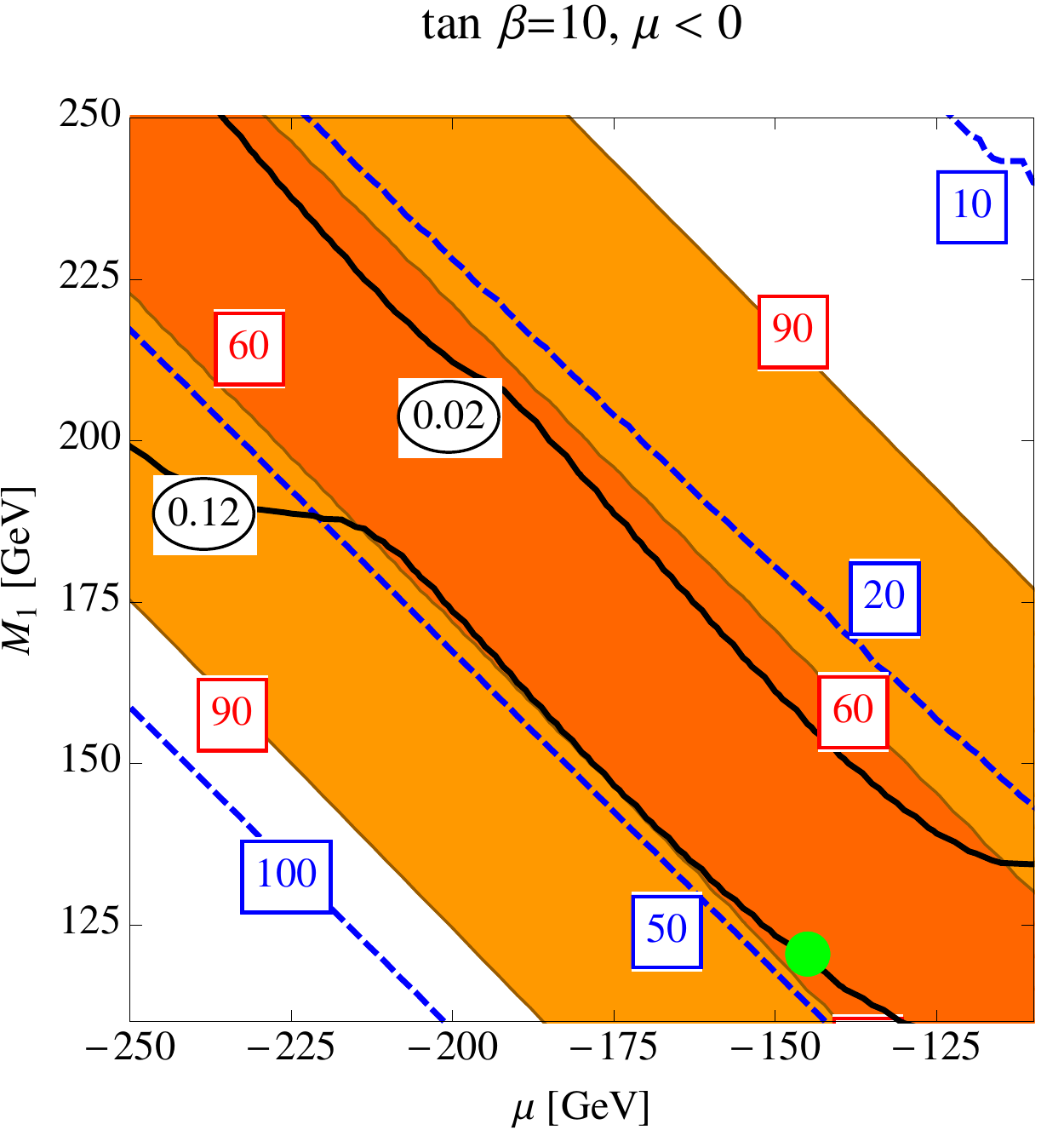}
\includegraphics[width=0.45\textwidth]{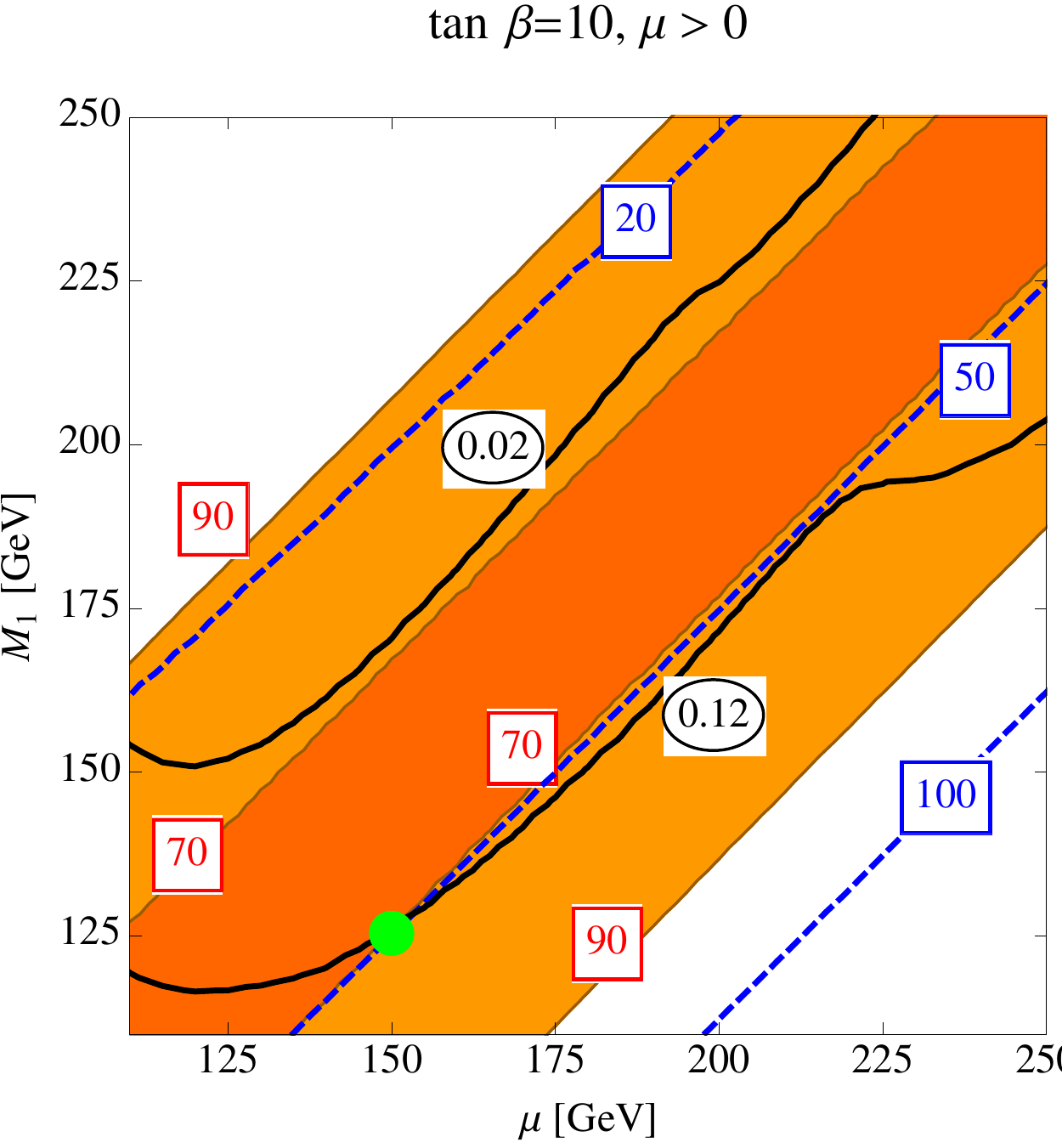}
\caption{Mass splitting and dark matter relic abundances are shown for bino-Higgsino admixtures. These plots assume all other sparticles have much larger masses $\sim 3 ~\rm TeV$. The mass splitting between the next-to-lightest neutralino ($\widetilde{\chi}_2^0$)  and the lightest neutralino ($\widetilde{\chi}_1^0$) measured in GeV are indicated with dashed blue lines. The orange bands display the mass splitting between $\widetilde{\chi}_3^0$ and $\widetilde{\chi}_1^0$. Note that between the innermost orange bands, the splitting is less than the mass of the $Z$ boson, forcing off-shell decays to the LSP. The dark, inner bands exemplify the minimal range of this mass splitting. The black lines show dark matter relic abundances, $\Omega h^2 = 0.12$ (in accord with current observations) and $\Omega h^2 = 0.02$, a permissible relic abundance assuming other dark matter particles are present.}
\label{fig:params1}
\end{figure} 

Although low (sub-TeV) mass Higgsino and bino dark matter are separately disfavored, if the bino mixes appreciably with the Higgsino, a viable relic dark matter candidate -- the `well-tempered' neutralino scenario -- can emerge. Consider the smooth transition from the case where $M_1 \ll |\mu|$ to that of $M_1 \gg |\mu|$ by simultaneously lowering $|\mu|$ and raising $M_1$ (moving along a diagonal line from the lower-outer edge to the upper-inner edge of the plots in Fig.~\ref{fig:params1}). As the mass of $|\mu|$ and $M_1$ get closer, more annihilation channels open to the bino LSP via an off-shell Higgs or chargino coannihilation, and the relic abundance decreases. Thus we expect that for a well-tempered bino-Higgsino, the correct relic abundance is achieved in parameter space where $M_1 \lesssim |\mu|$.

The required bino-Higgsino mass relation is made more explicit in Fig.~\ref{fig:params1}, where we overlay the relic abundance contours on top of the contours of inter-Higgsino splitting. We find that the Higgsino and bino mass parameters producing the correct dark matter abundance for a well-tempered bino-Higgsino can be approximated by
\begin{align}
M_1 \simeq |\mu| - 25 ~{\rm GeV}
\label{eq:massparapprox}
\end{align} 
in the limit that all sparticles besides the bino and Higgsino are heavy and with only mild dependence on $\tan{\beta}$ and the sign of $\mu$. The neutralino relic abundance is indicated in Fig.~\ref{fig:params1} by  black lines in each of the panels. The line marked with $\Omega h^2 = 0.12$ in an oval corresponds to the well-tempered parameter space yielding the observed dark matter abundance\footnote{In this study we calculate bino-Higgsino relic abundance with micrOMEGAs3 \cite{Belanger:2013oya}, and use a mass spectrum derived with Suspect2 \cite{Djouadi:2002ze} for values set at the low scale to avoid complications with large logs coming from the decoupling of the 
other states, we do not include radiative corrections to the masses of the neutralinos or charginos.}. 

Dark matter direct-detection experiments, such as CDMS, XENON, and LUX have placed some constraints on well-tempered parameter space Refs.~\cite{Aprile:2013doa,Akerib:2013tjd,Xiao:2014xyn,Agnese:2013rvf}. However, recent work in Ref.~\cite{Cheung:2012qy} has emphasized that near certain pieces of the well-tempered region, and especially for tan $\beta \leq 2$ and ${\rm sign} (M_1) \neq {\rm sign} (\mu)$, the LSP of the bino-Higgsino will have a vanishing coupling to the Higgs. Thus for these well-tempered regions of $M_1-\mu$ parameter space, spin-independent direct detection will be less-sensitive to a relic bino-Higgsino, making concomitant collider studies of the bino-Higgsino especially important for probing the entirety of MSSM dark matter possibilities. For example, the two points in green in the top left panel of Fig.~\ref{fig:params1} have $\sigma_{SI} \lesssim 10^{-45} {~\rm cm^2}$, where  $\sigma_{SI}$ is the spin-independent dark matter-nucleon scattering cross-section\footnote{We use output from micrOMEGAs3 \cite{Belanger:2013oya} to determine DM-nucleon scattering.}. The recent LUX result, which constrains a $m_\chi \sim 100 {~ \rm GeV}$ LSP to have $\sigma_{SI} < 2 \times 10^{-45} {~\rm cm^2}$ at 90$\%$ confidence, Ref.~\cite{Akerib:2013tjd}, does not exclude these points. In addition, if one allows the CP-odd neutral Higgs mass $m_A$ to be light, Ref.~\cite{Huang:2014xua}, yet heavy enough to avoid $A \rightarrow \tau \tau$ searches at the LHC, the plausible nucleon-scattering blind regions extend to more parameter space. Particularly, for small tan $\beta$ and $-\mu \sim M_1$, as studied in this paper, the bino-Higgsino would be entirely unconstrained by any planned direct detection study, so long as $m_A \sim 300 ~{\rm GeV}$. It is important to note that nucleon-scattering blind regions exhibit some electroweak fine-tuning. Indeed, it has long been appreciated that because well-tempered neutralino relic abundance is sensitive to small shifts in electroweakino and Higgsino mass parameters, there is fine-tuning associated with the simple requirement that mixed neutralinos freeze out with the correct dark matter relic abundance, Ref.~\cite{ArkaniHamed:2006mb}. For more lengthy discussions see Refs.~\cite{Perelstein:2011tg,Perelstein:2012qg,Cheung:2012qy}.  

Pushing past the well-tempered region, as $\mu$ is lowered closer to $M_1$, the lightest neutralino will annihilate more efficiently, and the total neutralino relic abundance will continue to drop, because the LSP will be more Higgsino-like. A line in parameter space that fits this description is $\Omega h^2 = 0.02$, shown on all of the plots in Fig.~\ref{fig:params1}. In most situations, this still lies in the shaded orange regions of parameter space, where all of the mass splittings of the lightest three neutralinos is less than $m_Z$, meaning this `well-forged' spectrum can also be found at colliders via decays of neutralinos to photons and dileptons, a topic explored at more length in Section \ref{sec:subSignal}.

\section{Branching Ratios and Cross-Sections for bino-Higgsinos}
\label{sec:subSignal}

In Section \ref{sec:fullspec} we showed that a wide swathe of parameter space for which the bino-Higgsino splittings are $O(20-70 {\rm ~ GeV})$ produces a fraction or the entire dark matter relic abundance. Past searches for light electroweakinos have focused on $pp\rightarrow \widetilde{\chi}^{\pm} \widetilde{\chi}^0_{2,3}$ in the context of a three lepton signal, where the bino-Higgsino splittings are greater than mass of $Z$ or $W^{\pm}$ boson, see Refs.~\cite{Gori:2013ala,Han:2013kza,Buckley:2013kua,Schwaller:2013baa,Baer:2014cua,Han:2014kaa,Low:2014cba,Baer:2014yta}.

To gain sensitivity to quasi-degenerate electroweakino spectra, another option is needed. One possibility is to search for electroweakinos produced in association with hard initial state radiation, looking in the final state $\slashed E_T + j + X$. As the additional radiation in the event can be used as a trigger, subsequent cuts can be relaxed and soft decay products from the decays among the electroweakinos can be picked out. In the extreme limit of electroweakino splittings $\ll \gev$, this technique becomes a mono-jet search, a standard dark matter collider signature~\cite{Birkedal:2004xn,Giudice:2010wb,Goodman:2010yf,Bai:2010hh,Goodman:2010ku,Fox:2011pm,Han:2013usa}. Though this technique is sensitive to smaller mass splittings, the need to produce hard initial state radiation in addition to the electroweakino pair reduces the cross section substantially, order $50$ ($p_{T,j} > 100\,\gev, 14\,\tev)$~\cite{Han:2014kaa} and signal rate becomes the limiting factor.

\subsection{Bino-Higgsinos in photon decays} 

Rather than rely on associated radiation to access compressed electroweakino spectra, we propose looking for pair production of bino-Higgsino electroweakinos in the final state $\ell^+\,\ell^- \gamma + \slashed E_T$. 

\begin{figure}[h!!]
\begin{center}
\includegraphics[width= 2.5 in]{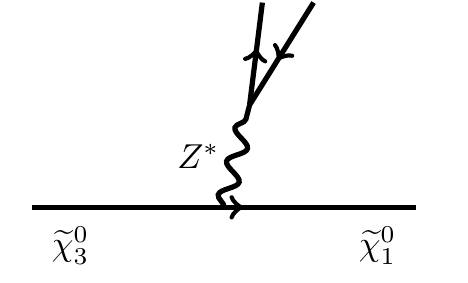}
\includegraphics[width= 2.5 in]{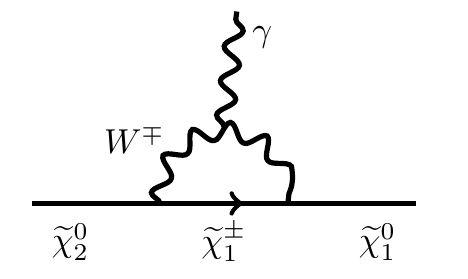}
\caption{Decays of $\widetilde{\chi}^0_3$ through a dilepton pair and $\widetilde{\chi}^0_2$ through a photon.}
\label{fig:ndecays}
\end{center}
\end{figure}

While there are several possible avenues for electroweakino pairs to arrive at this state, the strategy we advocate is best suited to pair production of heavy neutralinos which decay, one to $\ell^+\ell^- \widetilde{\chi}^0_1$ and the other to $\gamma +\widetilde{\chi}^0_1$. Neutralino decays to photons are often neglected, since the decay is a loop-level process, proceeding via a $W^{\pm}-$chargino loop. However, when the neutralino spectrum gets squeezed, the photon decay mode becomes competitive. Specifically, as the splitting among neutralinos shrinks below $m_Z$, neutralino decays through the $Z$ become three-body decays and are phase-space suppressed. Combined with the small branching fraction of the $Z$ to leptons -- the most clearly identifiable decay products -- it is certainly feasible that $BR(\widetilde{\chi}^0_{2,3} \to \gamma\,\widetilde{\chi}^0_1) \cong BR(\widetilde{\chi}^0_{2,3} \to Z^{*}(\ell^+ \ell^-) \widetilde{\chi}^0_1)$. We will make this relation among decay modes more concrete shortly.  One set of Feynman diagrams showing the $\widetilde{\chi}^0_{2,3} \to \ell^+ \ell^- \widetilde{\chi}^0_1$ and $\widetilde{\chi}^0_{2,3} \to \gamma \widetilde{\chi}^0_1$ decays are given in Fig.~\ref{fig:ndecays}.

Having specified the final state we intend to study, the viability and sensitivity of our search depends on i.) the rate of electroweakino (specifically neutralino) production, ii.) the branching fraction of the neutralino pairs into the $\ell^+\ell^-\gamma + \slashed E_T$ final state, and iii.) the size and kinematic characteristics of the SM backgrounds. The production cross section and branching fractions of neutralinos vary as we move in bino-Higgsino parameter space ($\mu, M_1, \tan{\beta}$) and will be addressed in turn in this section. We will study the SM backgrounds in more detail in Sec. \ref{sec:collid}.

\subsection{Production of bino-Higgsinos}
\label{sec:sigma}

\begin{figure}[h!!]
\begin{center}
\includegraphics[width=0.45\textwidth]{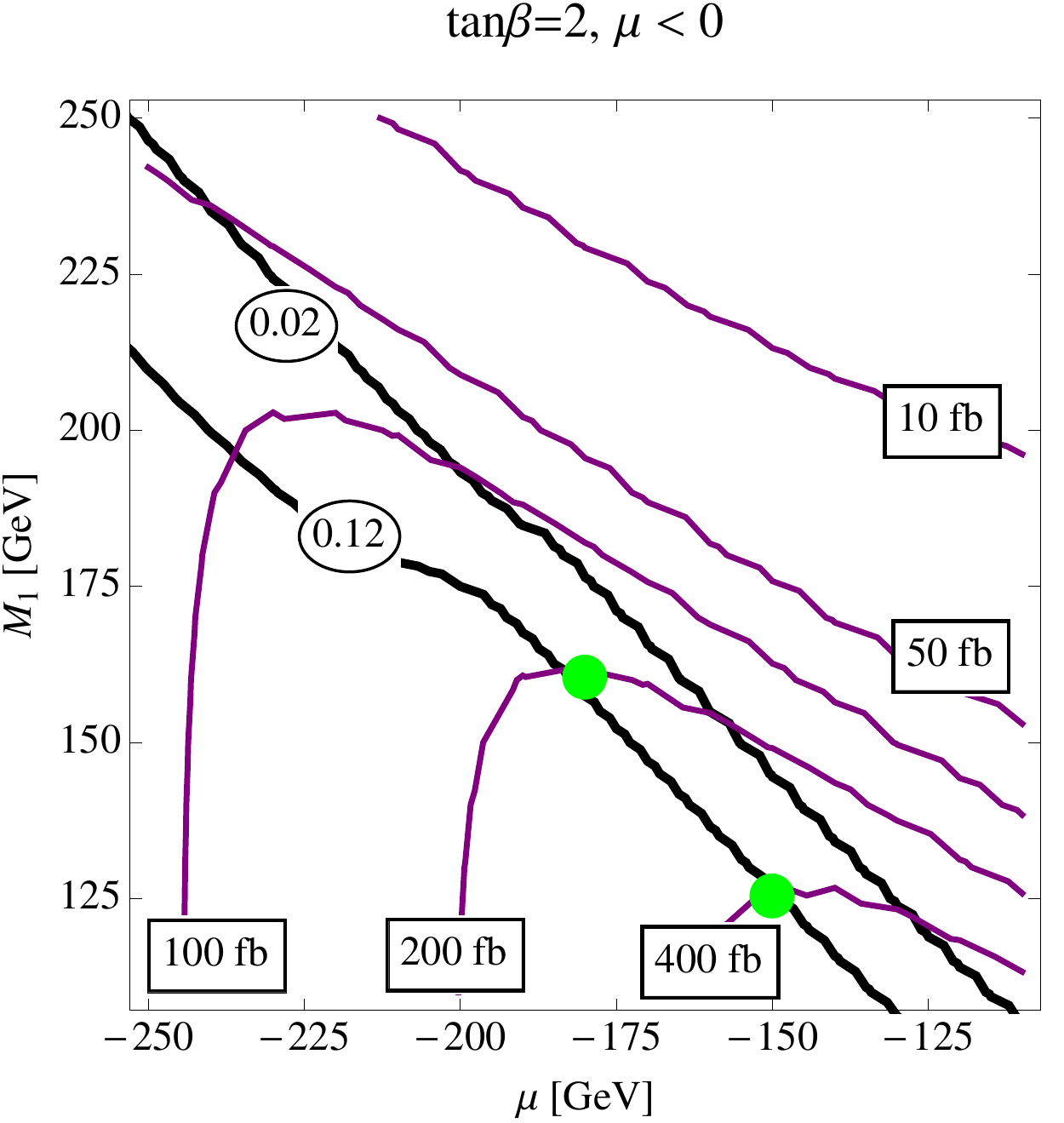}
\includegraphics[width=0.45\textwidth]{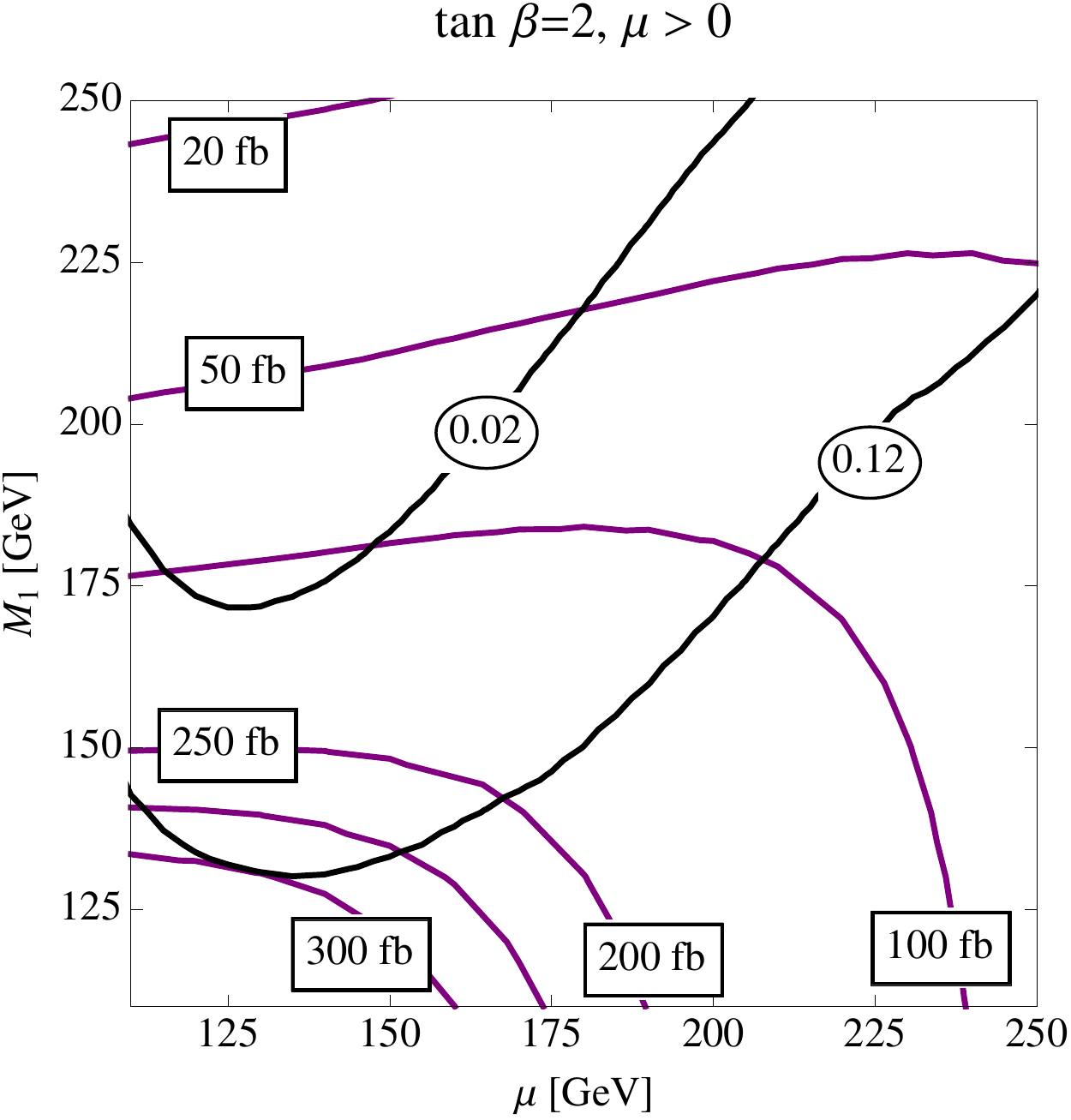}
\includegraphics[width=0.45\textwidth]{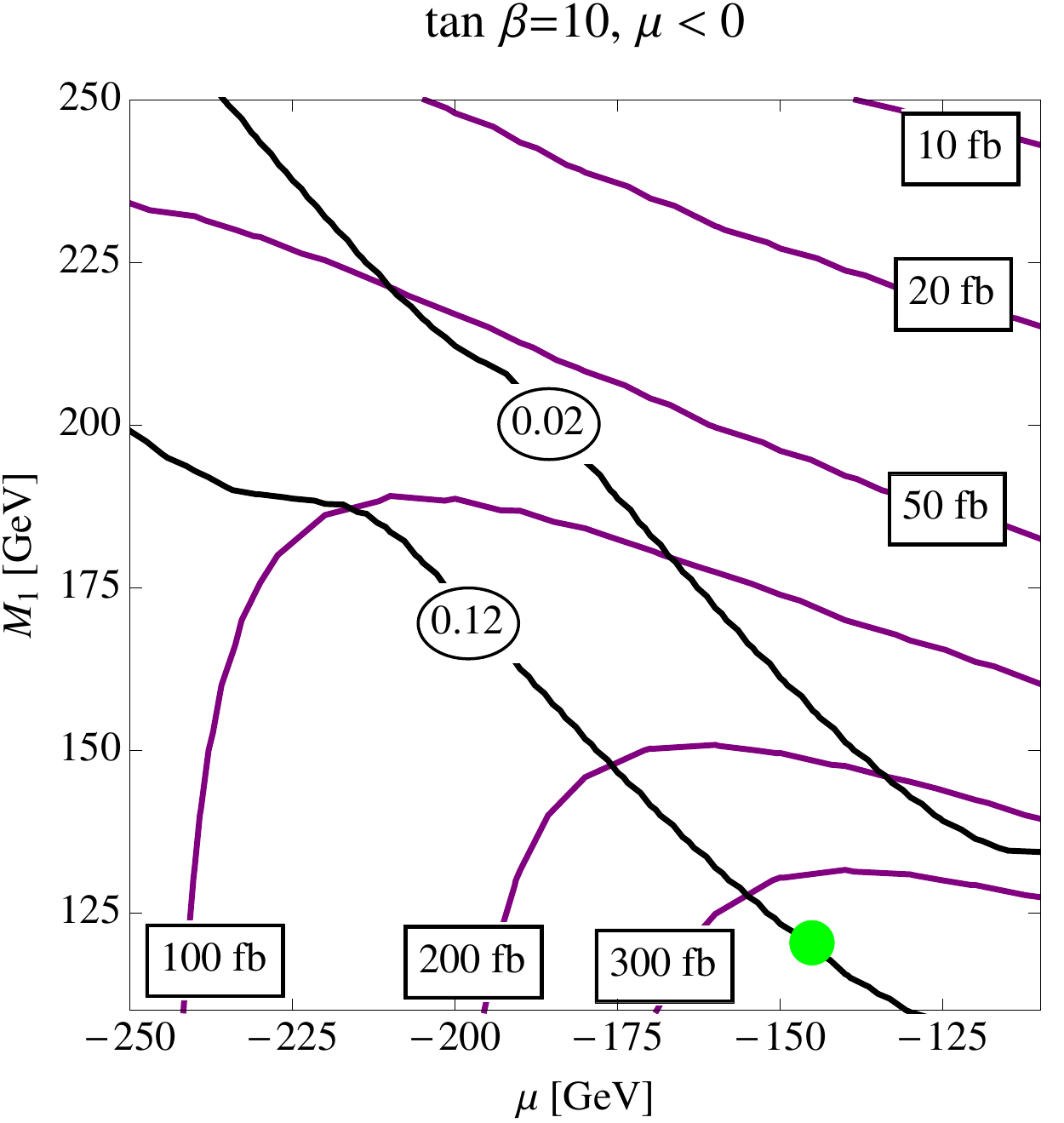}
\includegraphics[width=0.45\textwidth]{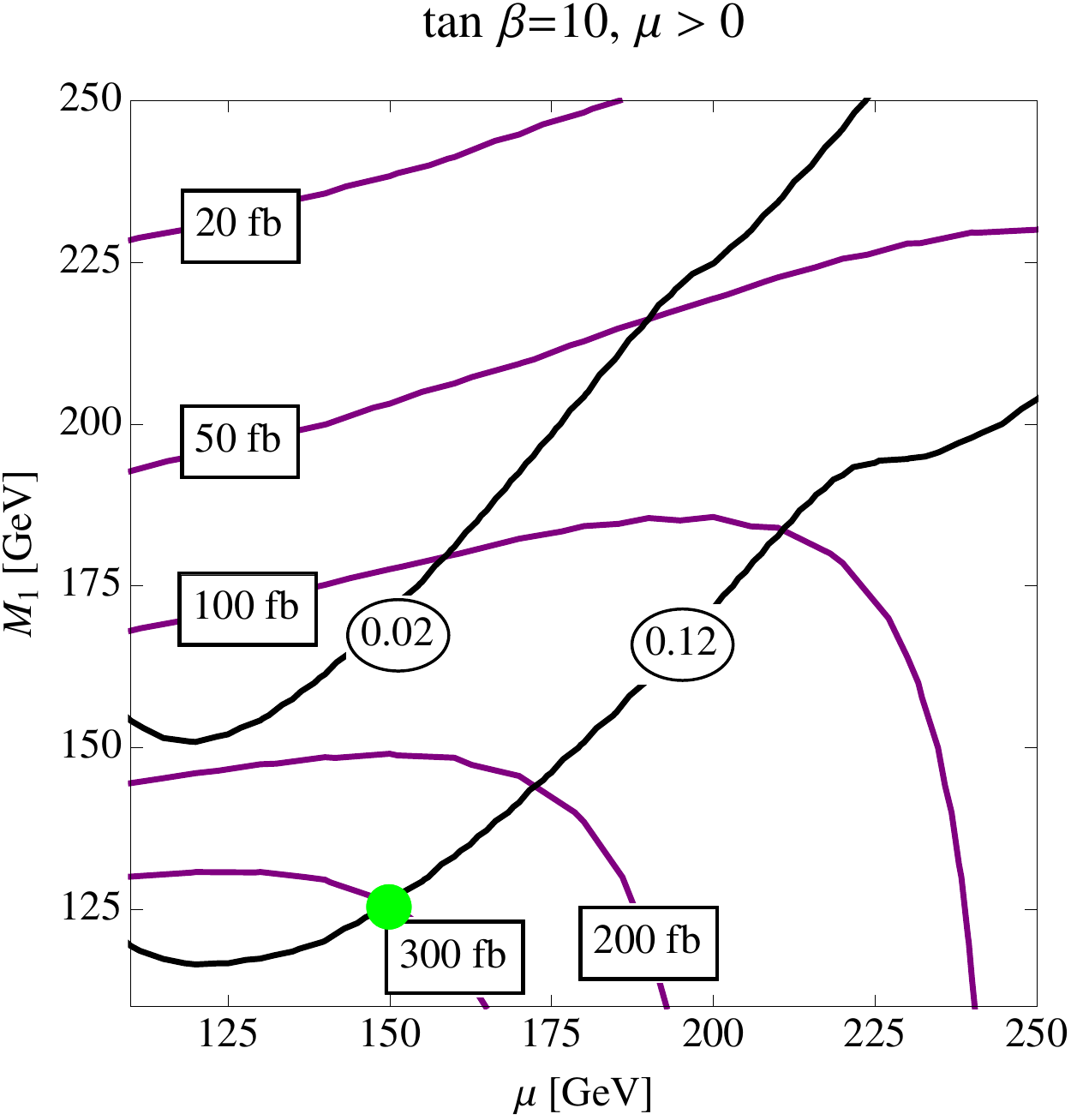}
\caption{Lines for the bino-Higgsino relic abundance and the cross section $pp \rightarrow \widetilde{\chi}_2^0 \widetilde{\chi}^0_3$ at the 14 TeV LHC are indicated with oval bubbles and rectangular bubbles, respectively. The bottom black line sets the relic abundance observed in our universe. The upper black like has a relic abundance of 0.02 which is allowable if there is another dark matter candidate. The green points in parameter space are studied in this paper for the signal $pp\rightarrow \widetilde{\chi}^0_2\widetilde{\chi}^0_3 \rightarrow \widetilde{\chi}^0_1\widetilde{\chi}^0_1 \ell^+ \ell^- \gamma$.}
\label{fig:crosssections}
\end{center}
\end{figure}

Turning first to the production, one element of the signal rate is how many electroweakino subprocesses contribute to our final state. Several different electroweakino pair-production modes are possible, i.e. $\widetilde{\chi}^{\pm}_1 \widetilde{\chi}^{\mp}_1,\, \widetilde{\chi}^0_2\,\widetilde{\chi}^0_2, \widetilde{\chi}^0_3\,\widetilde{\chi}^0_1$, etc., 
however as we will show later on, the mode driving the $\ell^+ \ell^- \gamma + \slashed E_T$ signal is $pp \to \widetilde{\chi}^0_2 \widetilde{\chi}^0_3$. In Fig.~\ref{fig:crosssections}, we plot the production cross-section of these heavier neutralinos, $pp \rightarrow \widetilde{\chi}^0_2\,\widetilde{\chi}^0_3$ as a function of $\mu$ and $M_1$ for $\tan \beta = 2,\,10$. The cross sections are largest when the neutralinos are lightest and decrease more slowly as $|\mu|$ is increased compared to increasing $M_1$.

Mixed bino-Higgsinos are produced through an s-channel $Z$ or $W^{\pm}$ boson. However, as the bino is inert under $W^{\pm}/Z$ interactions, the neutralino mass eigenstates are produced in proportion to their Higgsino fraction. In the mass range pertinent to LHC studies, the well-tempered line that quenches the observed relic abundance of dark matter has $M_1$ about 25 GeV less than $|\mu|$. In this case, the production cross section will be larger for the heavier neutralinos than the lightest neutralino, because $\widetilde{\chi}^0_2$ and $\widetilde{\chi}^0_3$ have larger Higgsino components than $\widetilde{\chi}^0_1$. This is illustrated in Fig.~\ref{fig:crosssections} where we see a sharp drop in the cross section when $|\mu| < M_1$ indicating a large bino component in $\widetilde{\chi}^0_2, \widetilde{\chi}^0_3$. One might expect $pp \to \widetilde{\chi}^0_2 \widetilde{\chi}^0_2, \widetilde{\chi}^0_3 \widetilde{\chi}^0_3$ to have a similar size cross section as $pp \to \widetilde{\chi}^0_2 \widetilde{\chi}^0_3$, however due to the fact that the two Higgsinos have opposite hypercharge, the $Z$ couplings to same-flavor neutralinos (i.e. $\widetilde{\chi}^0_i \widetilde{\chi}^0_i$) are highly suppressed compared to mixed flavor. 

\subsection{Branching fraction of bino-Higgsinos}
\label{sec:br}

\begin{figure}[h!!]
\begin{center}
\includegraphics[width=0.49\textwidth]{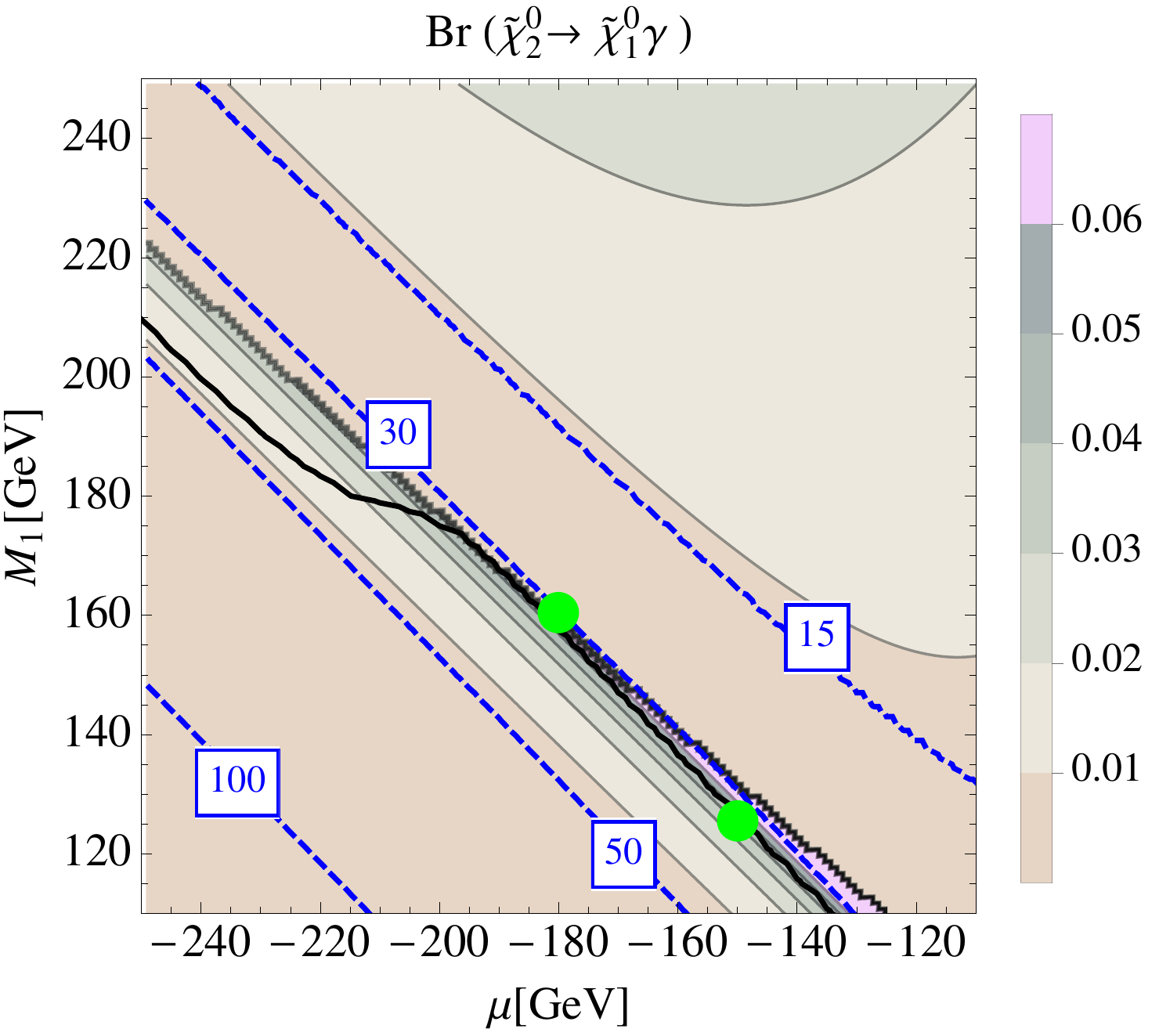}
\includegraphics[width=0.48\textwidth]{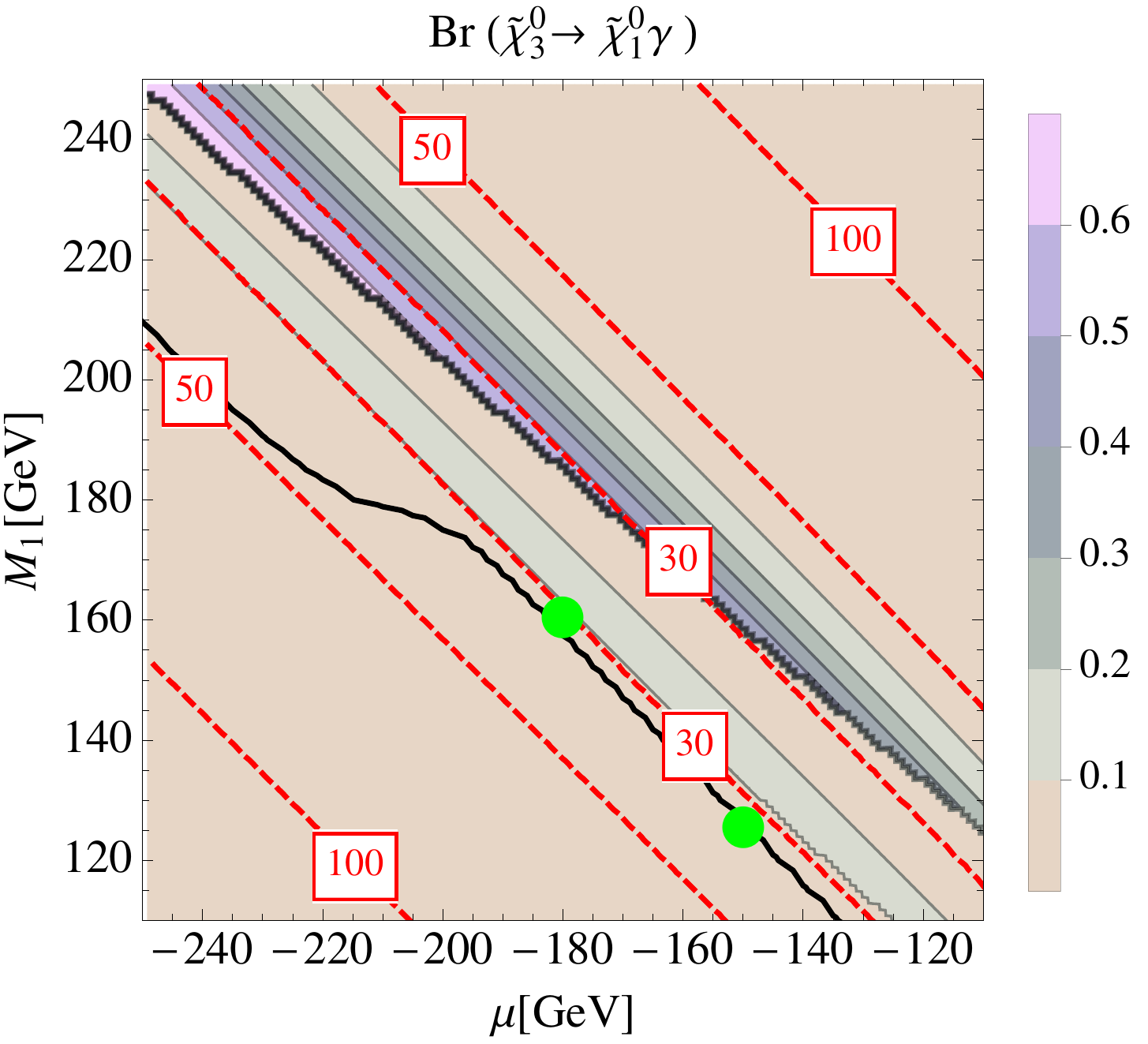} \\
\includegraphics[width=0.5\textwidth]{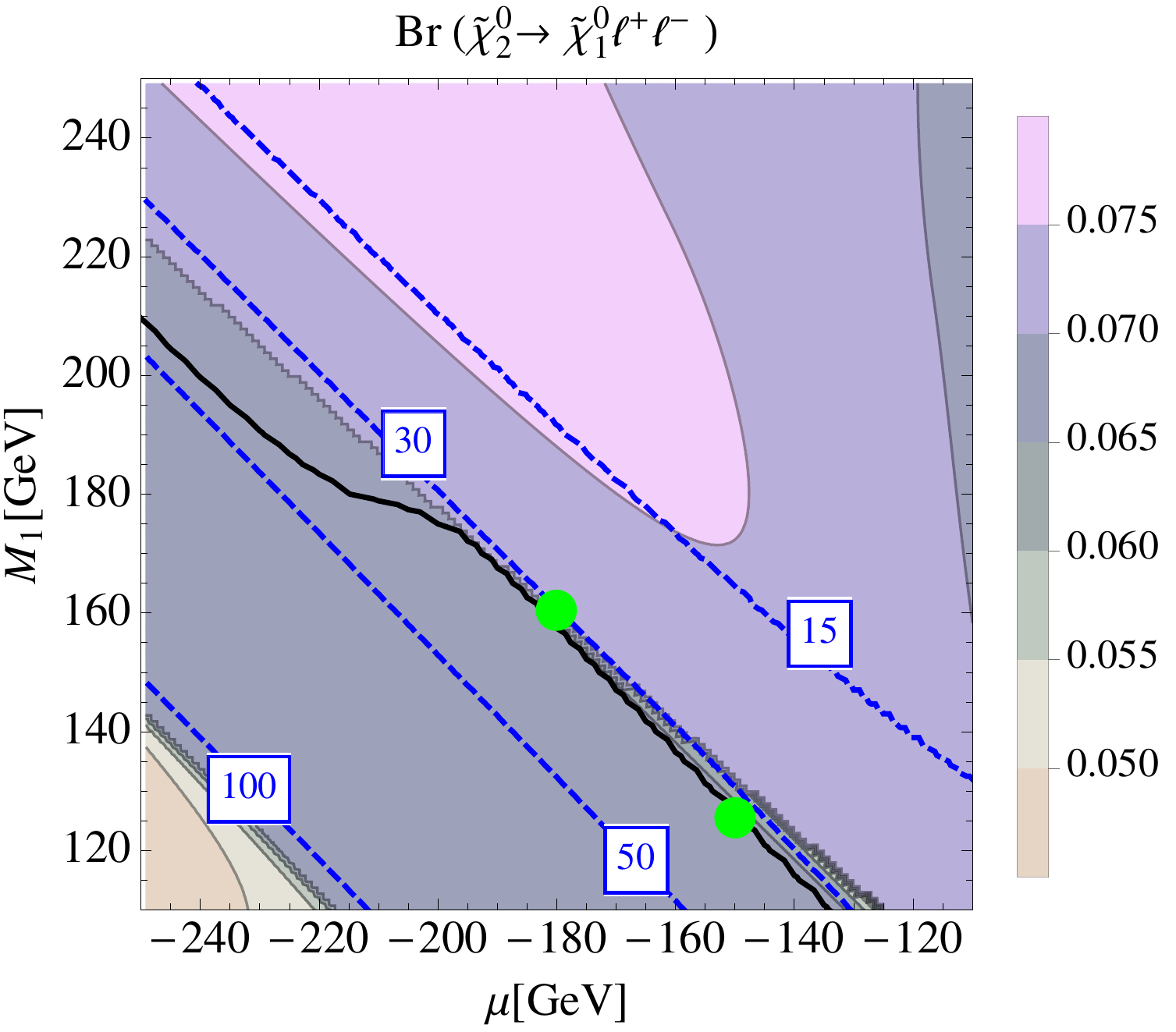}
\includegraphics[width=0.49\textwidth]{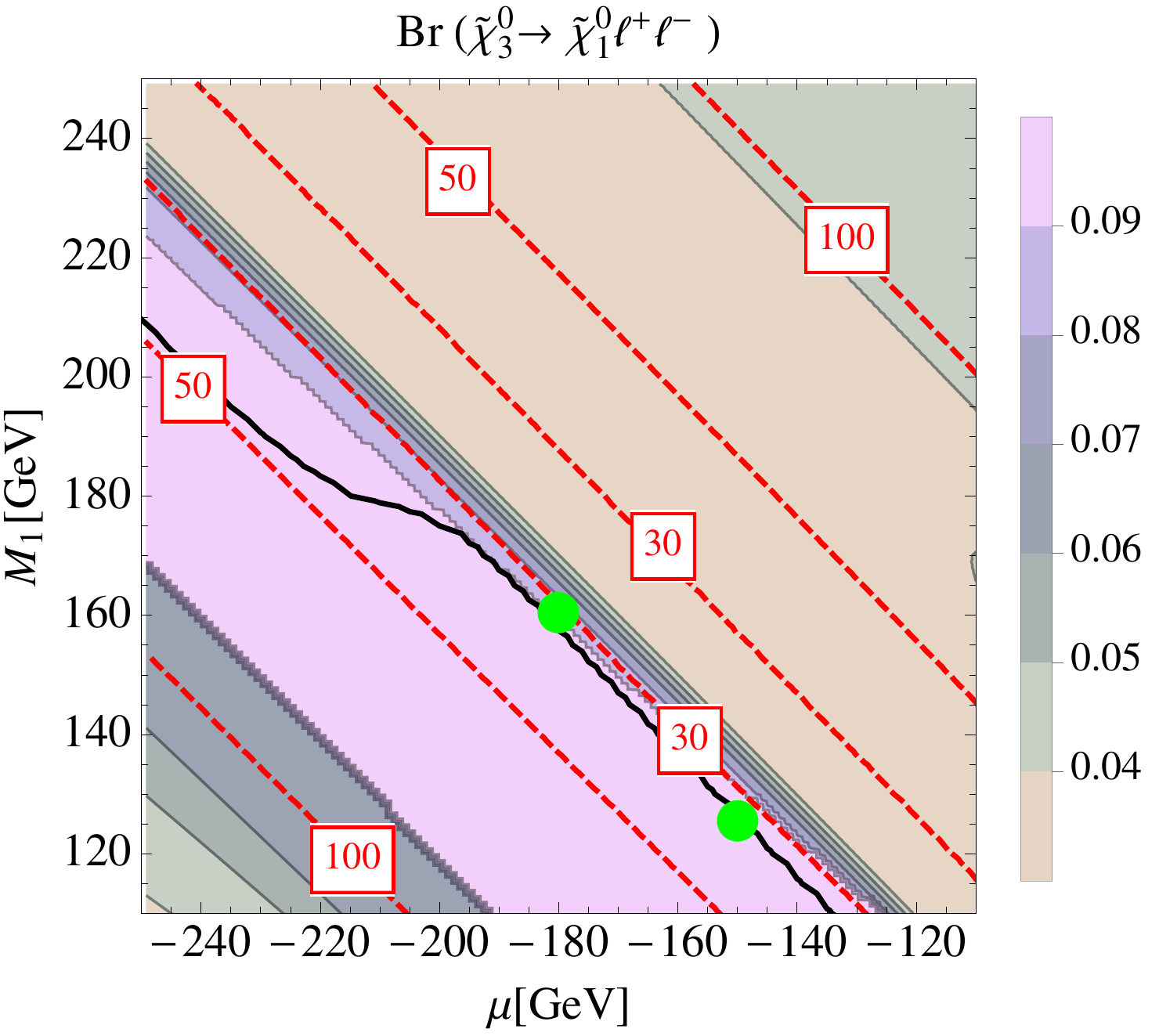}
\caption{The branching fraction for bino-Higgsinos decays to photons or dileptons and the LSP are shown for $\tan\beta=2$ and $\mu<0$. The black line is the well-tempered region indicating where bino-Higgsinos produce the observed relic abundance in our universe. The green points mark the benchmarks studied in section \ref{sec:collid}.}
\label{fig:thebR}
\end{center}
\end{figure}

The next ingredient is the branching fraction of $\widetilde{\chi}^0_3 \widetilde{\chi}^0_2$ into $\ell^+\ell^-\gamma + \slashed E_T$. Of the two decays we are envisioning, $\widetilde{\chi}^0_{2,3} \to \gamma +\widetilde{\chi}^0_1$ is the more exotic~\cite{Komatsu:1985he, Haber:1985fe,Gamberini:1986eg, Barbieri:1987pa, Haber:1988px,Baer:2002kv} and worth further scrutiny. The branching ratios $BR(\widetilde{\chi}^0_2 \to \gamma \widetilde{\chi}^0_1)$ and $BR(\widetilde{\chi}^0_3 \to \gamma \widetilde{\chi}^0_1)$ are shown in Fig.~\ref{fig:thebR} as a function of $\mu$ and $ M_1$   for $\tan{\beta}=2$. We have overlaid the mass splittings $m_{\widetilde{\chi}^0_3} - m_{\widetilde{\chi}^0_1}$ and $m_{\widetilde{\chi}^0_2} - m_{\widetilde{\chi}^0_1}$ on the branching ratio contours, as the splitting controls how suppressed the competing off-shell $Z$ decay modes are. The size of $BR(\widetilde{\chi}^0_{2,3} \to \widetilde{\chi}^0_1 \gamma)$ roughly follows the size of the mass splitting and peaks where $|\mu| \sim M_1$, though the transition is sharper. The sharpness of the transition is due to a level crossing of the $\widetilde{\chi}^0_2, \widetilde{\chi}^0_3$ eigenvalues. Specifically, as the diagonal elements of Eq.~(\ref{eq:mmatrix}) become degenerate, the mixing angles get large, suddenly altering the composition of the neutralinos. If a neutralino (either $\widetilde{\chi}^0_2$ or $\widetilde{\chi}^0_3$) inherits a large bino component, its $Z$ couplings all drop. Since the dominant mechanism of $\widetilde{\chi}^0_2, \widetilde{\chi}^0_3$ decay is via $Z$, when these couplings drop, the total width drops, and the branching ratio to photons -- which involves a different set of mixing parameters than the $Z$ modes -- jumps.

Combining the production and decay rates, we see that the  $\ell^+\ell^-\gamma+ \slashed E_T$ final state explored in this paper is well suited for, but not limited to, well-tempered neutralino parameter space. We now move on to the third factor in this mode's viability, the SM backgrounds, and suggest a set of collider analysis cuts to separate this background from the electroweakino signal.

\section{Compressed Electroweakinos from Photon + Dilepton at the LHC}
\label{sec:collid}

The collider final state we are interested in extracting from compressed electroweakinos is $\ell^+\ell^- + \gamma + \slashed{E}_T$. In the standard model, there are a number of processes which give rise to this final state. The dominant backgrounds for the electroweakino $\gamma + \ell^+\ell^-+ \slashed E_T$ signal are
\begin{equation}
\begin{aligned}
pp\rightarrow& \left. t\overline{t}~\gamma\right|_{\text{dilepton decay}} \\
pp\rightarrow& \left.\gamma^*/Z(\tau^+\tau^-)~\gamma\right|_{\text{dilepton decay}} \\
pp\rightarrow& \left.VV~\gamma\right|_{\text{dilepton decay}}
\label{eqn:backgrounds}
\end{aligned}
\end{equation}
where the photon is radiated from a charged particle in the initial or final state. In the $VV\gamma$ background, $V$ corresponds to all combinations of $W^{\pm}/Z/\gamma^*$, though in practice the dominant contribution comes from $W^+W^-\gamma$. The presence of missing energy, multiple electromagnetic objects, and little to no hadronic activity strongly limits what backgrounds can arise. There are other processes which can contribute to the $\ell^+\ell^-\gamma + \slashed E_T$ final state through object mis-reconstruction (fakes) or other realities of pileup and hadronic chaos in the LHC environment. We believe that, for the final state we are interested in, these environmental backgrounds are manageable. We will therefore ignore them for now, deferring more detailed comments until Sec. \ref{sec:discussion}.

To show that the $\ell^+\ell^-\gamma + \slashed E_T$ electroweakino final state can be effectively discriminated from these backgrounds, we pick four benchmark points from the well-forged and well-tempered parameter space. These points are marked as green dots in Figs.~\ref{fig:params1}, \ref{fig:crosssections}, and \ref{fig:thebR}. The points A and B both have a negative value for $\mu$ and $\tan\beta = 2$, which leads to small mass splittings between the neutralinos. Points C and D have $\tan\beta=10$, which creates larger mass splittings. A summary of these benchmark points is given in Table \ref{tab:Benchmarks}. It will be shown that the smaller mass splitting in points A and B not only leads to a higher branching ratio to photons, but also leads to more distinct kinematics than the larger splitting of points C and D.

\begin{table}[h!]
\begin{center}
\begin{tabular}{l| r r r r}
\hline
\hline
Benchmark points  & Point A & Point B & Point C & Point D \\
\hline
$\mu$ & -150 GeV & -180 GeV & -145 GeV  & 150 GeV\\
$M_1$ & 125 GeV & 160 GeV & 120 GeV  & 125 GeV\\
$\tan\beta$ & 2 & 2 & 10 & 10\\
\hline
$m_{\widetilde{\chi}^0_1}$ & 124.0 GeV & 157 GeV & 105 GeV & 103 GeV \\
$m_{\widetilde{\chi}^0_2}$ & 156.9 GeV & 186 GeV & 150 GeV & 153 GeV\\
$m_{\widetilde{\chi}^0_3}$ & 157.4 GeV & 188 GeV & 163 GeV & 173 GeV\\
\hline
$\sigma(pp\rightarrow \widetilde{\chi}^0_2 \widetilde{\chi}^0_3)$ & 394 fb & 200 fb & 345 fb & 287 fb \\
$BR(\widetilde{\chi}^0_2 \rightarrow \widetilde{\chi}^0_1 \gamma)$ & 0.0441 & 0.0028  & 0.0017 & 0.0014\\
$BR(\widetilde{\chi}^0_2 \rightarrow \widetilde{\chi}^0_1 \ell^+\ell^-)$ & 0.0671 & 0.0712 & 0.0702 & 0.0700\\
$BR(\widetilde{\chi}^0_3 \rightarrow \widetilde{\chi}^0_1 \gamma)$ & 0.0024 & 0.0767 & 0.0115 & 0.0102\\
$BR(\widetilde{\chi}^0_3 \rightarrow \widetilde{\chi}^0_1 \ell^+\ell^-)$ & 0.0714 & 0.0613 & 0.0447 & 0.0304\\
\hline
$\sigma(pp\rightarrow \widetilde{\chi}^0_2 \widetilde{\chi}^0_3 \rightarrow \gamma \ell^+ \ell^- \widetilde{\chi}^0_1 \widetilde{\chi}^0_1)$ & 1.297 fb & 1.125 fb & 0.279 fb & 0.205 fb\\
\hline
\hline
\end{tabular}
\caption{Values of interest for the four benchmark points highlighted in this analysis. These points are marked with green dots in Figs.~\ref{fig:params1}, \ref{fig:crosssections}, and \ref{fig:thebR} (A,B) . Points A and B have negative values for $\mu$ and $\tan\beta=2$, which leads to smaller mass splittings between the neutralinos. Points C and D have $\tan\beta=10$ which creates larger splittings. The larger mass splitting of points C and D leads not only to smaller branching ratios to photons, but also makes the signal kinematics more similar to the backgrounds.}
\label{tab:Benchmarks}
\end{center}
\end{table}

There are also electroweakino processes other than $pp \rightarrow \widetilde{\chi}^0_2 \widetilde{\chi}^0_3$ which generate a $\ell^+\ell^-\gamma\,+ \slashed E_T$ final state. For example:
\begin{align}
pp \rightarrow& \gamma \left(\widetilde{\chi}^+ \rightarrow \widetilde{\chi}^0_1 \ell^+ \nu_{\ell}\right) \left(\widetilde{\chi}^- \rightarrow \widetilde{\chi}^0_1 \ell^- \overline{\nu}_{\ell}\right),
\label{eqn:alt1}
\\
pp \rightarrow& \left(\widetilde{\chi}^0_2\rightarrow j j \widetilde{\chi}^0_1 \right) \left(\widetilde{\chi}^0_3 \rightarrow \gamma \widetilde{\chi}^0_2 \rightarrow \gamma \ell^+\ell^- \widetilde{\chi}^0_1 \right),  
\label{eqn:alt2}
\\
pp \rightarrow& \left(\widetilde{\chi}^+ \rightarrow \widetilde{\chi}^0_1 j j^{\prime} \right)  \left(\widetilde{\chi}^0_3 \rightarrow \gamma \widetilde{\chi}^0_2 \rightarrow \gamma \ell^+\ell^- \widetilde{\chi}^0_1 \right),
\label{eqn:alt3}
\\
pp \rightarrow& \gamma ~ \left(\widetilde{\chi}^+ \rightarrow \widetilde{\chi}^0_1 j j^{\prime} \right) \left(\widetilde{\chi}^0_{2,3} \rightarrow \ell^+ \ell^- \widetilde{\chi}^0_1\right).
\label{eqn:alt4}
\end{align}
We refer to the processes in Eq.~(\eqref{eqn:alt1}-\eqref{eqn:alt4}), which are explained in more detail in Appendix~\ref{app:processes}, as `alternative signals' because they have a different final state photon kinematic distribution than the dominant signal $pp\rightarrow \widetilde{\chi}^0_2 \widetilde{\chi}^0_3 \rightarrow \gamma \ell^+\ell^- \widetilde{\chi}^0_1 \widetilde{\chi}^0_1$, and are harder to distinguish from the SM background. For instance, the two chargino production in \eqref{eqn:alt1} has nearly the same collider morphology as the $WW\gamma$ background. These alternative signals are lumped together with the primary process, $pp \to \widetilde{\chi}^0_2\,\widetilde{\chi}^0_3$, to form the electroweakino signal in all of our simulations. \\

We conducted an analysis of these benchmark points and their backgrounds using Monte Carlo event generators to simulate LHC proton-proton collisions with a center of mass energy $\sqrt{s}=14$ TeV. To generate supersymmetric mass parameters for the signal events, we used the spectrum generated by SuSpect 2.43 \cite{Djouadi:2002ze} with the decays calculated by SUSY-HIT \cite{Djouadi:2006bz}. The resulting parameter card was used with PYTHIA 6.4~\cite{Sjostrand:2006za} to generate the events and perform subsequent showering, hadronization and decays. The background processes were generated with MG5@NCLO \cite{Alwall:2014hca}, again using PYTHIA 6.4 for showering and hadronization. To simulate collider acceptance in this analysis, we implemented jet clustering and required that partons pass angular cuts and $p_T$ thresholds as detailed below. For a more extended discussion of collider triggers and efficiencies, see Section \ref{sec:discussion}.

The final state involves exactly two leptons and one photon. We identify lepton candidates by requiring they have $|\eta| < 2.5$ and $p_T > 8$ GeV. Photon candidates also must have $|\eta| < 2.5$ and $p_T > 20$ GeV. We require the leptons to be isolated from each other, the photon, and jet candidates. For each lepton (photon) candidate we check the hadronic energy within a radius of $\Delta R < 0.4$. If the hadronic energy is greater than $5\%$ of the lepton (photon) energy, the lepton (photon) is added to the jet seeds. The jets were combined using Fastjet3 \cite{Cacciari:2011ma} with the anti-$k_T$ jet algorithm and jet radius of 0.5;  subsequently, we impose a minimum jet $p_T$ of 25 GeV and a rapidity of $|\eta| < 2.5$ on all jets. A final check removes any lepton or photon within $\Delta R < 0.4$ of a jet. The leptons are then sorted by their transverse momentum, defining the lepton with largest $p_T$ as $\ell_1$ and sub-leading $p_T$ as $\ell_2$. We then use the 8 TeV dilepton trigger defined as
\begin{equation}
p_{T,\ell_1} > 20 \text{ GeV} ~~\&~~ p_{T,\ell_2} > 8 \text{ GeV}.
\label{eqn:trigger1}
\end{equation}
For our analysis, we require that there are only two leptons, and that these two leptons need to be a same flavor opposite sign pair (SFOS). In addition, we require that all events have a single photon. The photon $p_T$ criteria is quite high
\begin{equation}
p_{T,\gamma} > 20 \text{ GeV},
\label{eqn:trigger2}
\end{equation}
 which helps to reduce soft photon backgrounds. For the rest of the analysis, we refer to equations \eqref{eqn:trigger1} and \eqref{eqn:trigger2} as the basic selection. 

\begin{figure}[h!]
\includegraphics[width=3in]{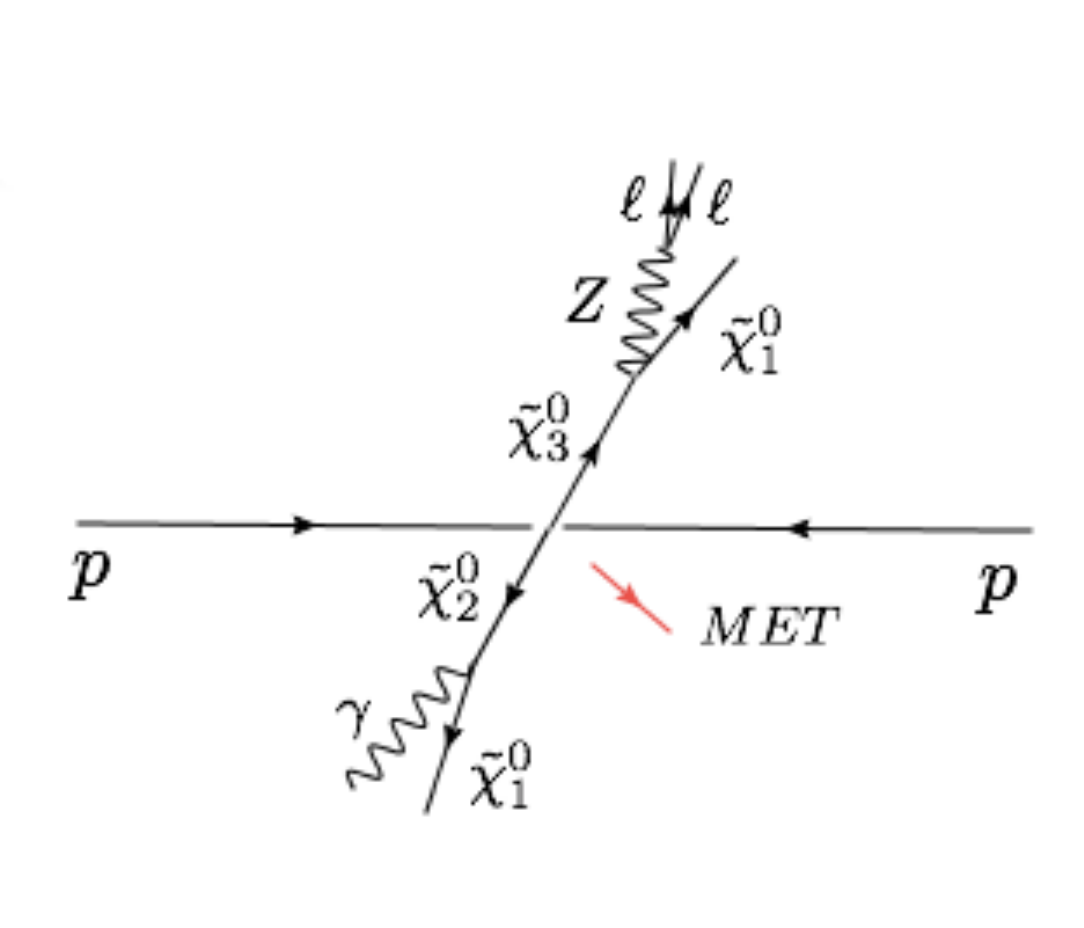}
\caption{An illustration of the signals characteristic kinematic features. The two leptons should be minimally separated, while the angle between the photon and the dilepton system should be large. The two $\chi^0_1$s are in nearly opposite directions leading to small amounts of missing energy. }
\label{fig:CartoonEvent}
\end{figure}

To further separate the signal from the background, we can make use of several kinematic features peculiar to the signal. Fig.~\ref{fig:CartoonEvent} shows an illustration of a possible event to help visualize the kinematic features.

\begin{itemize}
\item{No jets $p_T > 25\, \gev, ~|\eta| < 2.5$ in the event. The signal comes completely from electroweak production and therefore contains little hadronic activity. Meanwhile, backgrounds such as $t\bar t + \gamma$ are characterized by at least two jets and are strongly suppressed by this condition.}
\item{$|\Delta \phi_{\ell_1,\ell_2}| < \pi/2$, where $\Delta \phi_{\ell_1,\ell_2}$ is the azimuthal angle between the two leptons. In the signal, both leptons come from the decay of either $\chi^0_2$ or $\chi^0_3$ and tend to be close together. This is in contrast to the $VV\gamma$ and $t\overline{t}\gamma$ backgrounds where the leptons come from two separate $W$ bosons, or the $\gamma^*/Z(\tau^+\tau^-)+\gamma$ where the leptons come from two taus.  Thus, by placing a cut on the maximum $|\Delta \phi_{\ell_1,\ell_2}|$, we can remove a large fraction of the background without affecting the signal. The area normalized distributions for $|\Delta \phi_{\ell_1,\ell_2}|$ are shown in the first panel of Fig.~\ref{fig:CutSelection2}.
\begin{figure}[h!]\
\includegraphics[width= 3 in]{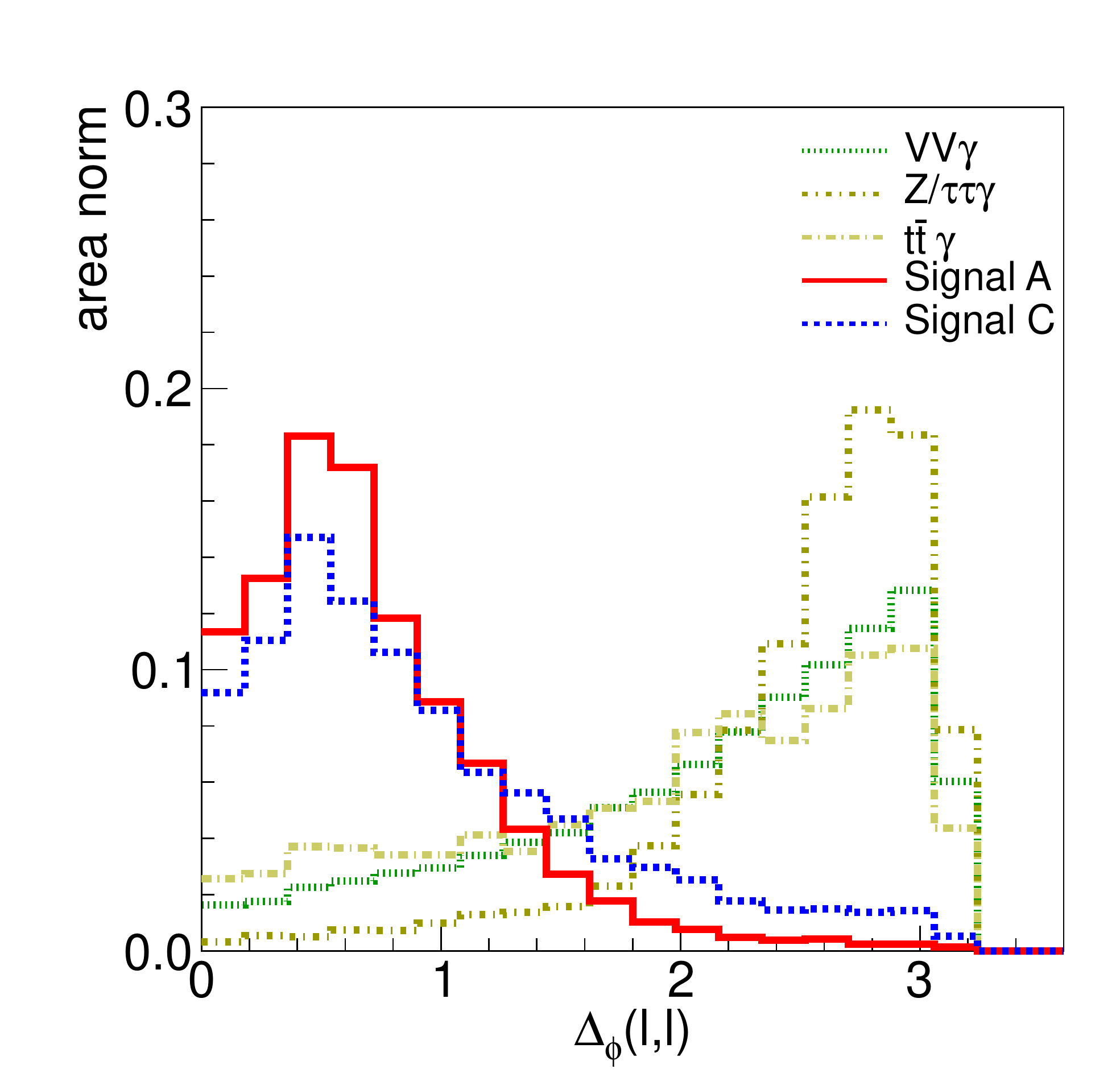}
\includegraphics[width= 3 in]{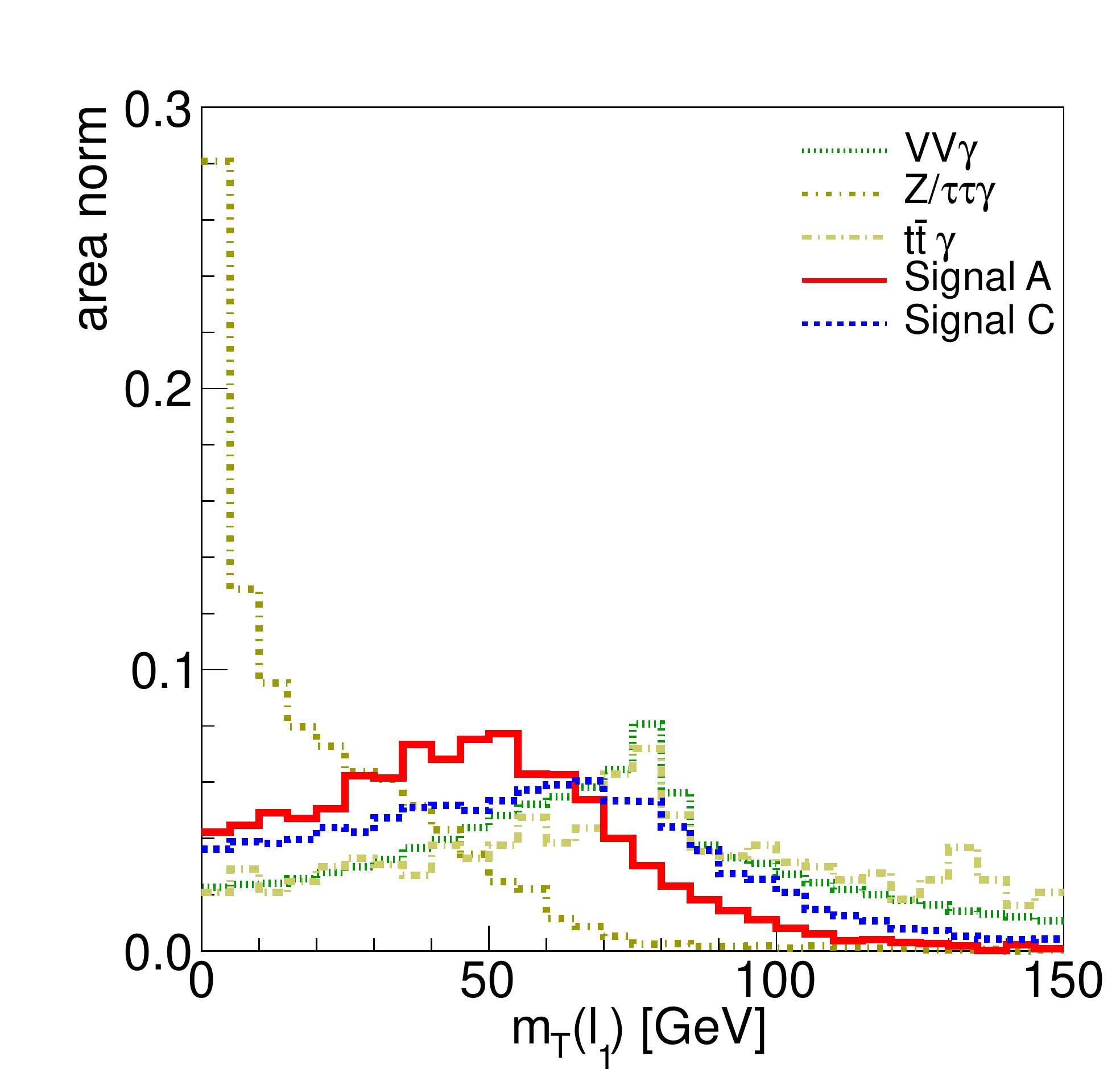}
\includegraphics[width= 3 in]{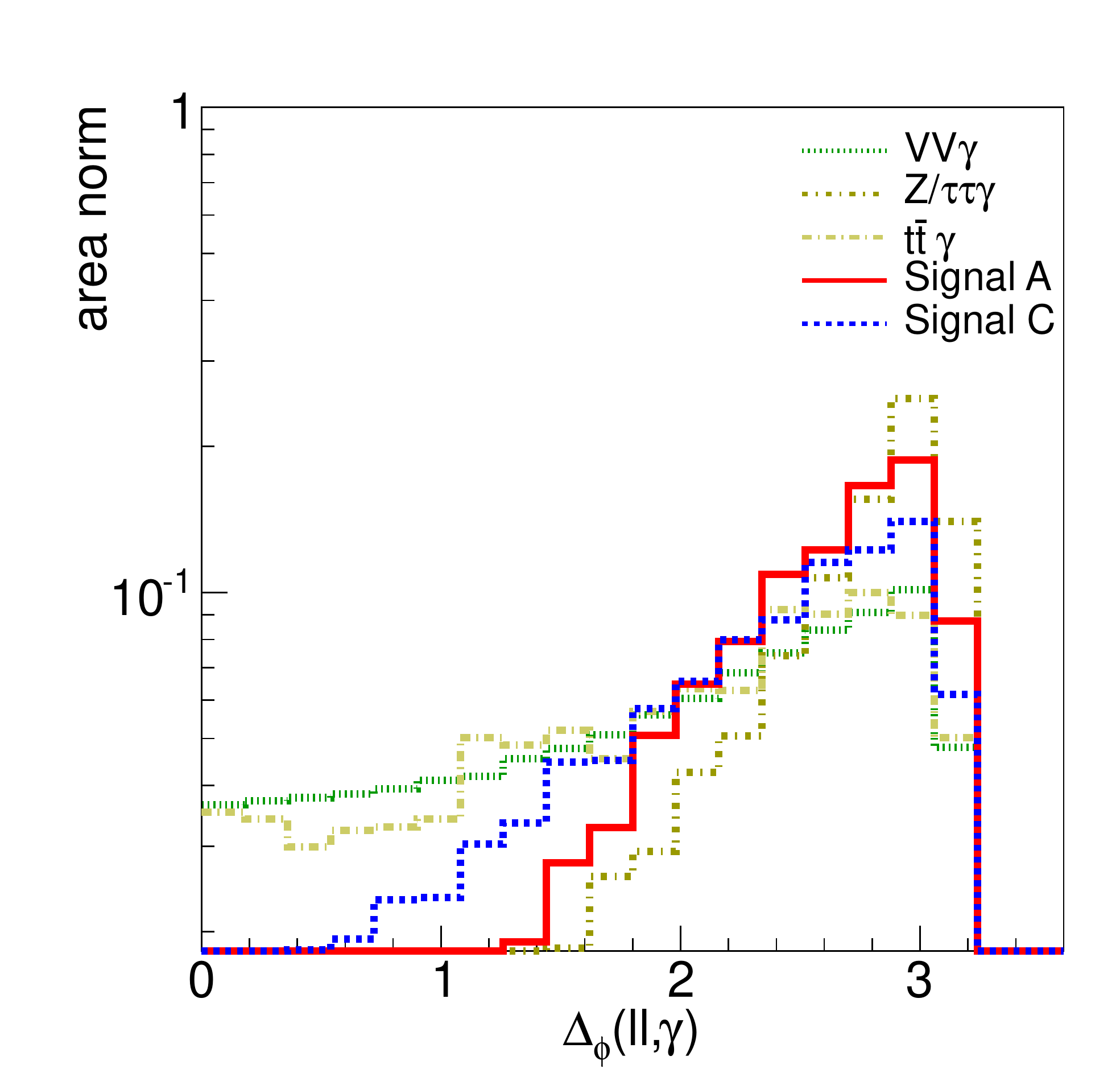}
\caption{Area normalized distributions of $\left|\Delta\phi_{\ell,\ell}\right|$, $m_T(\ell_1)$, and $\left|\Delta\phi_{\ell\ell,\gamma}\right|$ for events that have passed the trigger and the $0$ jet constraint. Point A has mass splitting of the neutralinos $\sim 25$ GeV while point C has splittings on the order of $50\,\gev$. The larger splitting causes all cuts to be less effective than the lower mass splitting case.
}
\label{fig:CutSelection2}
\end{figure}

}
\item{$10~\gev < m_T(\ell_i)\lesssim m_W$, where $m_T(\ell_i)$ is the transverse mass formed from either of the two leptons and the missing energy. 
A minimum threshold of $m_T(\ell_i, \slashed E_T) > 10\,\gev$ removes a large fraction of the $\gamma^*/Z(\tau^+\tau^-)+\gamma$ background without throwing away much of the signal.  An upper limit on $m_T(\ell_i,\slashed{E}_T) < m_W$ removes large portions of the $t\overline{t}+\gamma$ and $VV+\gamma$ backgrounds. The area-normalized distributions for $m_T(\ell_i, \slashed E_T)$ for the backgrounds of our benchmark signal points are shown below in the second panel of Fig.~\ref{fig:CutSelection2}. The $m_{T2}$ variable was also examined and found to provide good separation between signal and the $\gamma^*/Z(\tau^+\tau^-)+\gamma$. However, we found that using $m_T$ for both leptons individually provided better background discrimination than $m_{T2}$ for the other backgrounds. Note that, while in the preceding we have quoted a cut of $m_T(\ell_i,\slashed{E}_T) < m_W$, the actual value of this cut will be listed below, and will depend on the particulars of the parameter space point analyzed (i.e. do other necessary cuts already exclude $W$ boson containing backgrounds).
}
\item{$|\Delta \phi_{\ell\ell -\gamma} | > 1.0$, where $\Delta \phi_{\ell\ell -\gamma}$ is the azimuthal angle between the dilepton pair and the photon. In the signal the dilepton pair and the photon come from separate neutralino decays, $\chi^0_{2,3} \to \ell^+ \ell^- \chi^0_1,\, \chi^0_{3,2} \to \gamma \chi^0_1$ and therefore tend to be well separated in the detector. Photons that come from soft final state radiation, such as in the dominant $\gamma^*/Z(\tau^+\tau^-)+\gamma$ background, do not have this separation and are dominated by configurations where the photon is as close to one of the leptons as the isolation cuts allow.
}
\item{$m_{\ell\ell} \ll m_Z$. For the signal the maximum of this distribution is set by the inter-electroweakino splitting, while the background distributions is broad and peaked at $~\sim 50\,\gev$ ( $\sim 40\,\gev$ for $\gamma^*/Z(\tau^+\tau^-)+\gamma$). Therefore, by imposing a cut on the maximum allowed value of $m_{\ell\ell}$, we retain the signal while suppressing all backgrounds. The optimal $m_{\ell\ell}$ window depends on the signal point under consideration. For the sake of thoroughness, we note that we did not enforce a minimum invariant mass on $m_{\ell\ell}$.} 
\end{itemize}

In addition to these primary kinematic handles, we find several other variables that show small separation between the signal and the background. These include the photon $p_T$, the amount of missing energy, and the angles between the missing energy and the photon or dilepton system. Details of these cuts can be found in Appendix \ref{sec:ExtraCuts}. The two $\widetilde{\chi}^0_1$s are nearly back-to-back which yields a small amount of missing energy, and there is preferred orientation of the photon or dilepton relative to the $\slashed E_T$. This is in stark contrast to ISR-based searches~\cite{Schwaller:2013baa,Baer:2014cua,Han:2014kaa}, where the signal is characterized by large amounts of missing energy.

The actual numerical values that optimize the analysis vary from benchmark to benchmark. To determine the optimal set of cuts we scan over the possible lower and upper bound of the kinematic variables. At each step a simple significance, defined by $S/\sqrt{B}$, is calculated, where the signal cross section does not use the `alternative' signals. We keep the cut which maximizes this value as it leads to the smallest necessary integrated luminosity to achieve a significance of 5. After the optimal cut for each variable is found, the resulting significances are compared and the largest one is chosen. After each cut is chosen, the process starts over again keeping the previous cuts fixed. While it is likely that other optimization procedures would yield slightly different numbers, we believe our qualitative conclusions are robust.

\begin{centering}
\begin{table}[h!!]
\makebox[\textwidth]{
\begin{tabular}{l c c c c c | c  }
\hline
\hline
`small mass splitting' cuts&\multicolumn{5}{c}{Cross section [ab]} & \multicolumn{1}{c}{Significance}\\
\hline
Cut & Signal A & Signal B & $VV\gamma$ & $t\overline{t}\gamma$ & $Z/\tau\tau \gamma$ & S/B \\
\hline
0) Basic Selection & 281 & 169 & 5830 & 18900 & 24500 & 5.7$\times10^{-3}$ (3.4$\times10^{-3}$)  \\
1) $N_{jets}=0$ & 181 & 108 & 4820 & 1220 & 21400 & 6.6$\times10^{-3}$ (3.9$\times10^{-3}$)  \\
2) $\left|\Delta \phi_{\ell_1,\ell_2}\right| < 1.0$ & 118 & 79.5 & 580 & 201 & 567 & 8.8$\times10^{-2}$ (5.9$\times10^{-2}$) \\
3) $\left. \begin{matrix}15~\gev < m_T(\ell_2) < 50~\gev \\ m_T(\ell_1) < 60~\gev \end{matrix} \right\}$  & 52.4 & 38.2 & 93.3 & 32.8 & 92.2 & 0.24 (0.17)\\
4) $\left|\Delta\phi_{\ell\ell-\gamma}\right| > 1.45$  & 49.9 & 37.0 & 65.2 & 25.0 & 67.8 & 0.32 (0.23)\\
5) $30~\gev < p_{T,\gamma} < 100~\gev$  & 36.9 & 28.2 & 36.6 & 17.2 & 19.0 & 0.51 (0.39) \\
6) $\slashed{E}_T$ cuts & 26.8 & 20.2 & 24.6 & 3.90 & 0.00 & 0.94 (0.71) \\
7) $m_{\ell\ell} < 24~\gev$ & 23.3 & 19.3 & 9.29 & 0.00 & 0.00 & 2.5 (2.1) \\
\hline
\hline
\end{tabular}}
\caption{Cuts used to isolate the signal for benchmark points A and B. In the last column, the numbers not in parenthesis are for point A and the numbers in parenthesis are for point B.}
\label{tab:CutsForSig1}
\end{table}
\end{centering}

The benchmark points $A$ and $B$ have comparable splittings, which leads to very similar cuts. We therefore take the average of these cut values and define the `small mass splitting cuts'. The cut values and resulting significances are summarized below in Table \ref{tab:CutsForSig1}, where the signal cross sections now include the `alternative signals' of equations \eqref{eqn:alt1}-\eqref{eqn:alt4}. From these cuts we estimate that Point $A$ could be discovered with an integrated luminosity of 430 fb$^{-1}$ and Point B could be discovered with 620 fb$^{-1}$ of data.

Similarly, benchmark points $C$ and $D$ have comparable mass splittings so their cuts are averaged for the `large mass splitting cuts', which are shown in Table \ref{tab:CutsForSig2prime}. The benchmark points $C$ and $D$ have smaller initial cross sections, but the kinematics are also more similar to the backgrounds which makes the cuts less effective. We estimate that point $C$ will be take 4300 fb$^{-1}$ of integrated luminosity to discover, while point $D$ will take 1900 fb$^{-1}$. The required luminosities are large, but within the scope of a high-luminosity LHC run. 

\begin{centering}
\begin{table}[h!]
\makebox[\textwidth]{
\begin{tabular}{l c c c c c | c }
\hline
\hline
`large mass splitting' cuts&\multicolumn{5}{c}{Cross section [ab]} & \multicolumn{1}{c}{Significance}\\
\hline
Cut & Signal C & Signal D & $VV\gamma$ & $t\overline{t}\gamma$ & $Z/\tau\tau \gamma$ & S/B \\
\hline
0) Basic Selection & 256 & 411 & 5830 & 18900 & 24500 & 5.2$\times10^{-3}$ (8.3$\times10^{-3}$) \\
1) $N_{jets}=0$ & 157 & 227 & 4820 & 1220 & 21400 & 5.7$\times10^{-3}$ (8.3$\times10^{-3}$) \\
2) $\left|\Delta \phi_{\ell_1,\ell_2}\right| < 1.05$ & 68.3 & 109 & 618 & 208 & 608 & 4.8$\times10^{-2}$ (7.6$\times10^{-2}$) \\
3) $\left. \begin{matrix}10~\gev < m_T(\ell_1) < 100~\gev \\  10~\gev <m_T(\ell_2) < 95~\gev \end{matrix} \right\}$  & 47.9 & 72.2 & 389 & 127 & 117 & 7.5$\times10^{-2}$ (0.11)\\
4) $8~\gev < \slashed{E}_T < 95~\gev$ & 45.8 & 69.4 & 375 & 116 & 84.1 & 7.9$\times10^{-2}$ (0.12) \\
5) $m_{\ell\ell} < 39~\gev$ & 42.8 & 64.0 & 228 & 35.9 & 51.5 & 0.14 (0.20) \\
\hline
\hline
\end{tabular}}
\caption{Cuts used to isolate the signal for benchmark points C and D. In the last column, the numbers not in parenthesis are for point C and the numbers in parenthesis are for point D.}
\label{tab:CutsForSig2prime}
\end{table}
\end{centering}

\begin{figure}[h!]\
\includegraphics[width=0.46\textwidth]{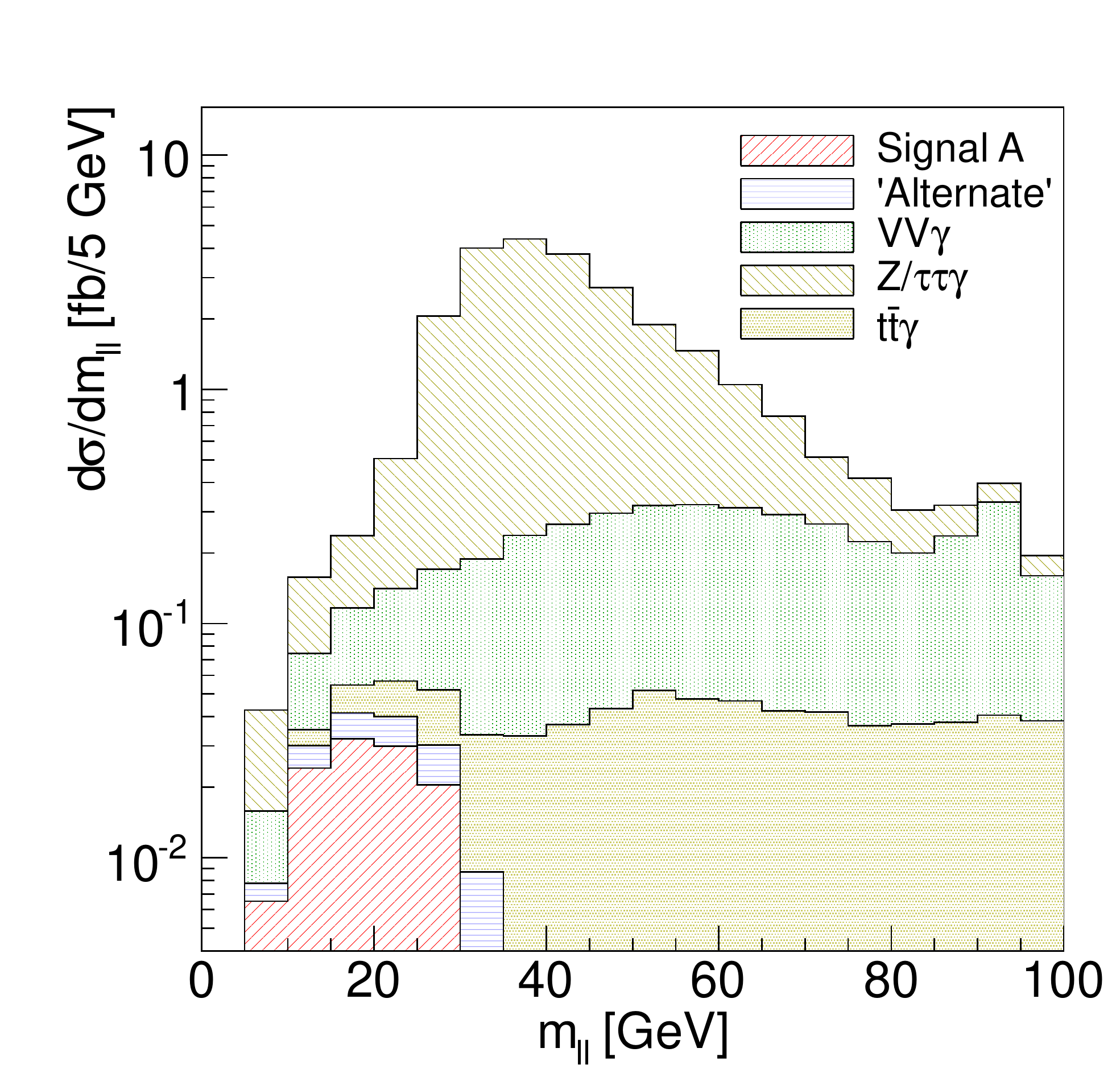}
\includegraphics[width=0.46\textwidth]{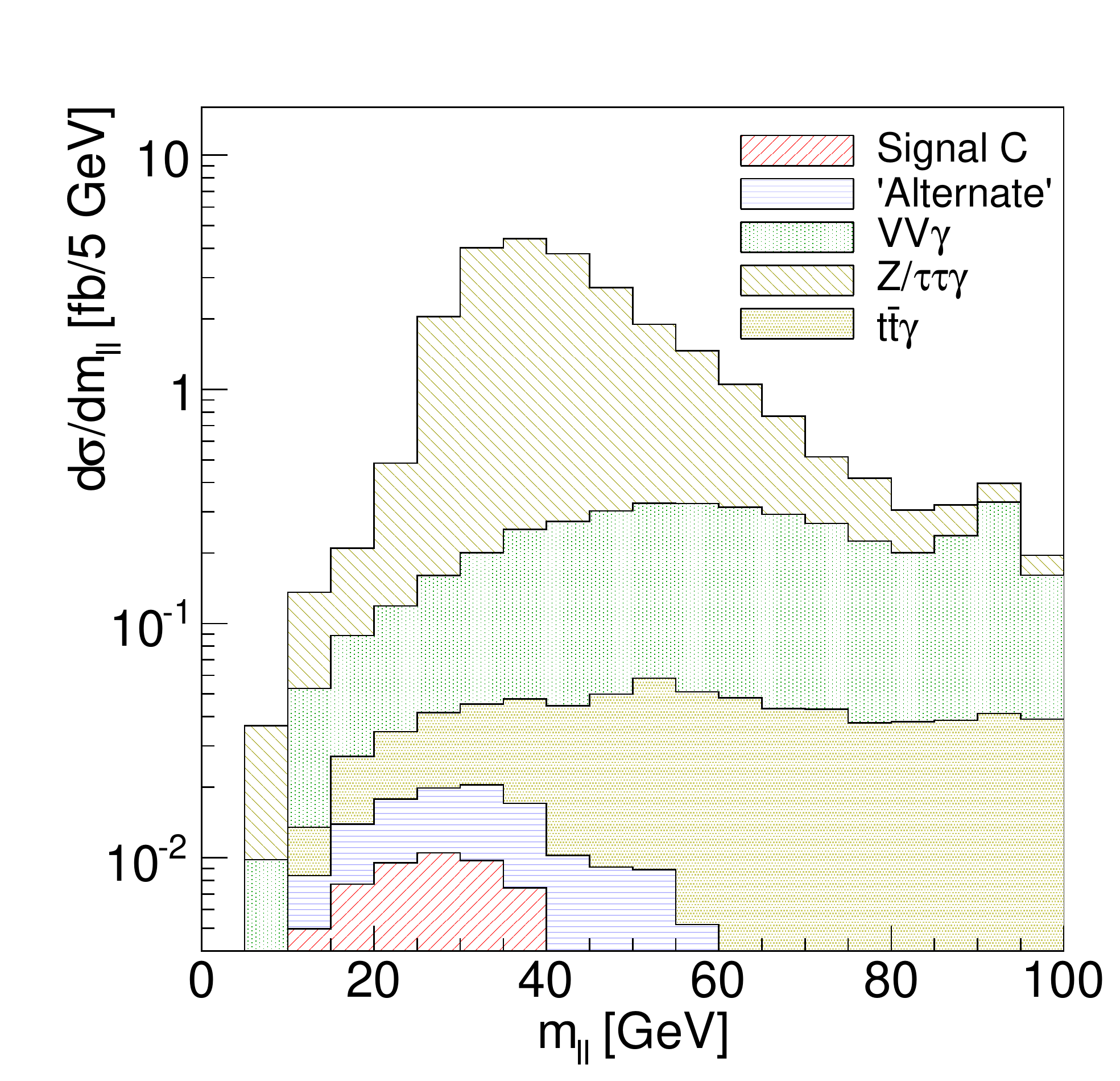}
\caption{Differential cross section of events passing the trigger and with 0 jets, but before applying any other cuts. The events in each bin are the sum of signal plus SM background contributions. The left (right) panel is for benchmark A (C). The red hatched region is the neutralino signal while the blue hatched is the extra `alternative' methods of achieving the same final state using electroweakinos.}
\label{fig:TriggerPass}
\end{figure}

We have shown that the $\ell^+\ell^-\gamma+\slashed{E}_T$ signal is more effective at the lower mass splittings of points $A$ and $B$ than it is for points $C$ and $D$. A large reason for this is the value of $m_{\ell\ell}$ which is determined by $m_{\widetilde{\chi}^0_{3,2}} - m_{\widetilde{\chi}^0_1}$. In Fig. \ref{fig:TriggerPass}, we plot the $m_{\ell\ell}$ distributions for points A and C. The red hashed regions are the signals examined in this paper and the blue region are the `alternative signals'. The small mass differences in point A leads to an $m_{\ell\ell}$ peak which is at lower values, which significantly helps reduce the $\gamma^*/Z(\tau^+\tau^-)+\gamma$ background. One then expects that the efficiency of this signal should get even better for lower mass splittings. However, as the splitting is decreased much more than the $\sim30~\gev$ observed in points A and B, the leptons become too soft to trigger on efficiently. We therefore expect that the smallest mass splitting, $\text{min}(m_{\widetilde{\chi}^0_{2}}-m_{\widetilde{\chi}^0_1}, m_{\widetilde{\chi}^0_{3}}-m_{\widetilde{\chi}^0_1})$, that this signal can be used for is $\sim25~\gev$. The regions of parameter space for this can be found in Fig. \ref{fig:params1}. 

The benchmark points C and D are harder to find with this signal, especially using only the process $pp\rightarrow \widetilde{\chi}^0_3\widetilde{\chi}^0_2 \rightarrow \ell^+ \ell^- + \gamma + \slashed{E}_T$. These points, which have a larger mass splitting (though $m_{\widetilde{\chi}^0_{3,2} }- m_{\widetilde{\chi}^0_1}$ are still below $m_Z$), suffer from a drop in the branching ratio to the photon as well as the effectiveness of the $m_{\ell\ell}$ cut. However, there is hope. When the difference in mass increases between $m_{\widetilde{\chi}^0_{2,3}}-m_{\widetilde{\chi}^0_1}$, the mass difference between $m_{\widetilde{\chi}^0_3} -m_{\widetilde{\chi}^0_2}$ also increases. This opens up the possibility of cascade decays such as those shown in Eqs. \eqref{eqn:alt2},\eqref{eqn:alt3}. For example, the main difference between points C and D is the difference in mass between $\widetilde{\chi}^0_2$ and $\widetilde{\chi}^0_3$: $13~\gev$ for point C and $23~\gev$ for point D. For point D, the $\widetilde{\chi}^0_3 - \widetilde{\chi}^0_2$ splitting is large enough that the photon and leptons from the decay
\begin{equation}
\widetilde{\chi}^0_3 \rightarrow \widetilde{\chi}^0_2 + (\gamma, \ell^+\ell^-)
\end{equation}
to be triggered. However, the geometry of the decays changes when the process goes through a cascade, and the photon and dilepton system are no longer back-to-back. The change in topology renders the $\left|\Delta \phi_{\ell\ell - \gamma}\right|$ ineffective, so we do not use include it in the `large mass splitting cuts' of Table \ref{tab:CutsForSig2prime}. The lepton separation $\left|\Delta\phi_{\ell_1,\ell_2}\right|$, the $m_T$ of the leptons, and $m_{\ell\ell}$ are still useful cuts, since the lepton properties are still constrained by the inter-electroweakino splitting. Using the `large mass splitting' set of cuts,  we can extend the region where this search method is effective to regions where all of the splittings are less than $m_Z$ and $\widetilde{\chi}^0_2$ and $\widetilde{\chi}^0_3$ are split by around $20-40~\gev$.  Applying some of the `small mass splitting' cuts, such as $|\Delta \phi_{\ell\ell-\gamma}|$ or $\slashed E_T$, to scenarios C and D does result in better $S/B$ than achieved in Table \ref{tab:CutsForSig2prime}, but the signal cross section drops so low that a much higher luminosity is needed to achieve a significance of $5$.

\section{Other Backgrounds}
\label{sec:discussion}

The analysis we have presented above neglects several details, which we discuss in more detail here. First, only physics backgrounds have been included; environmental backgrounds such as object misidentification (fakes) and overlapping partonic collisions have been neglected. Our analysis relies on the multiple electromagnetic objects, two leptons and a photon, which reduces the likelihood that our signal can be faked by multi jet processes. However, our analysis also relies on fairly soft leptons -- the subleading lepton $p_T$ cut is $8\,\gev$ -- and softer leptons are more easily faked by jets.

 Being more quantitative, one fake background comes from $W^{\pm}(\ell\nu) + \gamma + \jets$, where one of the jets is mistaken for a lepton. After basic cuts, the lowest order cross section at 14 TeV for $pp \to W(\ell\nu) + \gamma + \jet$ is $12.5\,\pb$. Randomly selecting one of the jets in the event to be treated as a second lepton then passing the `fake' $\ell^+\ell^-+\gamma+j$ evens through our `small-splitting scenario' analysis cuts, we find an efficiency of $0.12\%$. The net contribution of this fake process to the background is then the product of the signal rate, the analysis cut efficiency, and the rate for a jet to fake a lepton $\epsilon_{j\to \ell}$, which we take to be $p_T$ independent and fixed at $0.01\%$. This value is the most conservative rate quoted (for $p_{T,j} = 10\,\gev$) in the study in Ref.~\cite{Curtin:2013zua} (based on $7\,\tev$) multiplied by $1.3$ to account for the fact that we are simulating events at $14\,\tev$. The result is $\sigma(pp \to W(\ell\nu) + \gamma + j)_{fake} = 1.5\,\text{ab}$, which is small compared to both the other backgrounds and our benchmark signals.

In most supersymmetry searches the missing energy is large, so pure QCD backgrounds are not an issue. Our signal does not have large missing energy, so we have to consider a wider set of fakes.  One example is $pp \to \gamma + \jets$ with two jets faking leptons. Generating $\gamma + jj$ events with MG5@NCLO (including $\gamma \bar b b$), treating the two parton-level jets as leptons, and imposing all non-$\slashed E_T$ cuts, we find the rate to be $\sim 5\,\epsilon^2_{j\to\ell}\,\text{nb}$. To pass our signal, these events still need to acquire some $\slashed E_T$. Small amounts of $\slashed E_T$ are easy to acquire in the busy LHC environment from pileup or other soft interactions/decays. We estimate the faction of events with $\slashed E_T > 10\,\gev$ by the fraction of minimum bias events passing this threshold~\cite{ATLAS-CONF-2011-072,Chatrchyan:2011tn}, $\sim 10\%$\footnote{These estimates are based off of $7\,\tev$ data. At $14\,\tev$, higher pileup could make this fraction higher}. Including a factor of $0.5$ to crudely incorporate an efficiency to pass the `small-splitting' angle-related $\slashed E_T$ cuts and plugging in the value for $\epsilon_{j\to \ell}$, the result is $2.5\,\text{ab}$. As with $\sigma(W(\ell\nu) + \gamma + j)_{fake}$, this rate is subdominant to the irreducible background. A more accurate estimate would require overlaying minimum bias events on top of fake $\gamma + \jets$ events and treating the combination as single events. Such detailed treatment is beyond the scope of this paper.

A second environmental background worth mentioning is double parton scattering (DPS), two independent partonic collisions within the same initial proton pair. This background was brought up in Ref.~\cite{Han:2014kaa} in the context of electroweakino searches and found to be small. However, Ref.~\cite{Han:2014kaa} studied electroweakinos produced in association with a hard ISR jet, a qualitatively different kinematic region than we are studying here. Nevertheless, we believe the DPS background to be safely negligible because of the odd assortment of final state particles that our analysis employs. Specifically, while a $3\ell+\nu$ final state can be faked by the combination of  $pp \to W(\ell\nu)$ and a low-mass Drell-Yan event, there is no simple secondary process that can be combined with $W$ production to make $\ell^+\ell^- + \gamma + \slashed E_T$. Similarly, $pp \to Z(\nu\bar{\nu})$ is a useless ingredient because it provides no net $\slashed E_T$. One possible DPS candidate is $pp \to \ell^+\ell^-\gamma$ (Drell-Yan plus a photon emission) combined with $pp \to Z(\nu\bar{\nu})+j$. The cross section for $pp \to \ell^+\ell^-\gamma$ with basic cuts is $\sim 20\,\pb$, however after imposing all lepton and photon-based analysis cuts (but neglecting and $\slashed E_T$-based cuts) the rate drops to $72\,\fb$. The cut most responsible for suppressing $pp \to \ell^+ \ell^- \gamma$ is $|\Delta\phi_{\ell_1,\ell_2}| < 1.05$, since the leptons from $pp \to \ell^+ \ell^-\gamma$ are preferentially produced back-to-back. Combining the $pp \to \ell^+\ell^-\gamma$ rate with the cross section for $pp \to Z(\nu\bar{\nu})+j \sim 10\,\text{nb}$, and using the DPS estimation reviewed in Ref.~\cite{Berger:2009cm}, we find:
\begin{equation}
\sigma_{DPS}(pp \to (\ell^+\ell^-\gamma) + (Z(\nu\bar{\nu})j) ) = (72\,\fb)\times\frac{10\,\text{nb}}{(\sigma_{eff} = 12\,\text{mb})} \ll 1\,\text{ab}.
\end{equation}
This source of DPS background is orders of magnitude too small to impact our signal, even allowing for $O(1)$ variation in $\sigma_{eff}$ or the individual cross sections. 

While our study of environmental backgrounds has not been exhaustive, the low rates exhibited here give us confidence that our estimates based on physics backgrounds alone in Sec.~\ref{sec:collid} are reasonable.

Another place where our analysis has been optimistic is our use of $8\,\tev$ LHC lepton trigger thresholds. Once the LHC ramps up to 14 TeV, the increasingly chaotic environment may necessitate raising these thresholds. Higher thresholds hurt our analyses since our signal tends to have a softer lepton spectrum than the background. To quantify how increased thresholds affect the sensitivity, we have redone the previously presented analyses with lepton thresholds pushed to $30\,\gev$ for the leading lepton and $10\,\gev$ for the subleading lepton. With the higher thresholds, benchmark $A$ ($B$) requires $1400\,\fb^{-1}\, (2100\,\fb^{-1})$, roughly three times the value at lower threshold. The drop in significance is motivation for the 14 TeV LHC experiments to keep the lepton trigger thresholds as low as possible. The loss in significance may be offset somewhat by diversifying the search to include ISR, i.e.$pp \to \widetilde{\chi}^0_2\,\widetilde{\chi}^0_3 + j$, as the recoil of the electroweakinos off the initial jet is inherited by their decay products and can lead to higher trigger efficiency. This signal diversification is not free, however, since the background for $\ell^+\ell^-\gamma + \slashed E_T + j$ is large. A devoted study is needed to determine the ideal mixture of zero and one (or more) jet channels.  
 
Another place where our study could be improved is the modeling of the significance; we used a simple cut-based $S/\sqrt B$ measure to quantify the sensitivity. More sophisticated, multi-variate approaches can likely take additional advantage of the shape differences between the electroweakino signals and the SM backgrounds. Finally, all signal and background numbers have been computed using leading order cross sections. The $K$ factors for the signal and dominant backgrounds are similar and somewhat larger than $1$~\cite{Beenakker:1999xh, Kramer:2012bx,Hamberg:1990np,Melnikov:2011ta}. Simply slapping on these factors, $S/\sqrt B$ will shift slightly. However, our study focuses on a peculiar corner of phase space and it is possible that higher-order effects in this region are different than in the overall cross section.

\section{Conclusions}
\label{sec:conc}

In this work we have presented an alternate search channel for electroweakinos based on the final state $\ell^+\ell^-\gamma + \slashed E_T$. This final state comes about from a variety of electroweakino sources, but the signal we find most easily captured is $pp \to \widetilde{\chi}^0_2 \widetilde{\chi}^0_3$, where one of the heavier neutralinos decays to a lepton pair and the other to a photon and LSP.  The radiative decay mode $\widetilde{\chi}^0_{2,3} \to \gamma \widetilde{\chi}^0_1$ is usually ignored since it typically has a small branching fraction. However, when the electroweakino spectrum is compressed, more conventional electroweakino decay modes become suppressed and the $\gamma + \widetilde{\chi}^0_1$ mode can be competitive and even dominant. The parameter space where the electroweakino spectrum is compressed overlaps significantly with the so-called `well-tempered'  region, i.e. where admixtures of bino and Higgsino or bino and wino can act as dark matter. The lack of strong LHC bounds on compressed electroweakino spectrum, combined with the potential connection to dark matter makes seeking out new electroweakino search strategies a must.

Focusing on bino-Higgsino admixtures, we mapped out how quantities like the mass splitting, branching ratios, and relic abundance depend on the supersymmetry inputs. After identifying and studying the parameter space of interest, we presented our search strategy. By exploiting kinematic features of the signal such as low dilepton invariant mass, low hadronic activity, and small azimuthal separation between the leptons we were able to reduce the SM backgrounds ($VV\gamma, \gamma^*/Z(\tau^+\tau^-) + \gamma$ and $t\bar t\,\gamma$) enormously. This strategy is viable in any bino-Higgsino scenarios where the heavier neutralinos $\widetilde{\chi}^0_{2,3}$ are heavier than the LSP by $O(25-70\,\gev)$; if the splitting is smaller than $25\,\gev$, the final state particles are too soft to trigger on efficiently, while if the splitting is large enough that $\widetilde{\chi}^0_{2,3}$ can decay to an on-shell $Z$, the photon branching fraction plummets. Signal events with smaller $m_{\ell\ell}$ are easier to distinguish from the background, so our search performs best when $m_{\widetilde{\chi}^0_{2,2}} - m_{\widetilde{\chi}^0_1}$ is close to the lower threshold. Translated into supersymmetry parameters, the $\ell^+\ell^-\gamma + \slashed E_T$ search is best suited to $M_1 \lesssim |\mu|$, with $\mu < 0$ and small $\tan{\beta}$. As an example, we find neutralinos with spectrum set by $M_1 = 125\,\gev, \mu = -150\,\gev, \tan{\beta} = 2$ can be discovered with our technique with $430\,\fb^{-1}$. The amount of required luminosity increases as the overall mass scale of the electroweakinos is raised or as the splitting between $\widetilde{\chi}^0_{2,3}$ and the LSP grows. Once we increase the value of $\tan\beta$ the splitting increases and our signal becomes more difficult to differentiate from the background and we need luminosities at the ab$^{-1}$ level.

The search strategy we have demonstrated for the well-tempered bino-Higgsino could be applied to other dark matter frameworks. One simple application is to other neutralino mixtures, such as bino-wino, however it can also be applied to any mixtures of a light fermion singlet and fermion $SU(2)$ doublets with hypercharge $1/2$ or to $SU(2)$ charged scalar or vector dark matter with electroweak scale masses.  Indeed, for any dark matter state with couplings so small it would overclose the universe for an electroweak scale mass (i.e. the bino), if the annihilation rate for this state is increased via mixing with other heavy states that transform non-trivially under $SU(2)$ (like a pair of Higgsinos), the mass splittings between the singlet and heavier states can be detected via decays to photons and off-shell Z bosons. More generally and for the same reasons, the collider final state of $MET+\gamma+\ell^+ +\ell^-$ proposed in this article can be applied to any $\cal{O}(\text{200 ~GeV})$ relic dark matter whose freeze-out is dictated by couplings to electroweak gauge bosons.

For bino-Higgsino mixtures, the region where our search works best is exactly where direct detection searches struggle, since for low $\tan{\beta}$ and $\mu <0$ the couplings of the LSP to the Higgs boson are vanishingly small and the prospects for direct detection experiments are not great. It is important to stress again that searches for electroweakinos are not limited by the energy of the collision but by the luminosity, therefore a possible upgrade in the luminosity of the LHC could be key to be able to discover these blind spots.

\subsection*{Acknowledgments} We would like to thank Raffaele D'Agnolo for discussion.  The work of AD was partially supported by the National Science Foundation under Grant No. PHY-1215979, and the work of AM was partially supported by the National Science Foundation under Grant No. PHY-1417118.

\appendix
\section{Additional Photon + Dilepton Production Processes}
\label{app:processes}
In this appendix we detail alternate electroweakino dilepton + photon production modes. 

The process
\begin{equation}
pp\rightarrow \gamma \left(\widetilde{\chi}^+ \rightarrow \widetilde{\chi}^0_1 \ell^+ \nu_{\ell}\right) \left(\widetilde{\chi}^- \rightarrow \widetilde{\chi}^0_1 \ell^- \overline{\nu}_{\ell}\right).
\label{eqn:alt2a}
\end{equation}
involves the 2 to 3 body production of a photon and two charginos, each decaying leptonically. This process contains the same final state particles as the neutralino production considered. There are also collider processes which fall into the photon + dilepton category once we allow for extra final state products that are either soft or missed by the detector. The processes
\begin{equation}
\begin{aligned}
pp \rightarrow& \left(\widetilde{\chi}^0_2\rightarrow j j \widetilde{\chi}^0_1 \right) \left(\widetilde{\chi}^0_3 \rightarrow \gamma \widetilde{\chi}^0_2 \rightarrow \gamma \ell^+\ell^- \widetilde{\chi}^0_1 \right) \text{ and} \\
pp \rightarrow& \left(\widetilde{\chi}^+ \rightarrow \widetilde{\chi}^0_1 j j^{\prime} \right)  \left(\widetilde{\chi}^0_3 \rightarrow \gamma \widetilde{\chi}^0_2 \rightarrow \gamma \ell^+\ell^- \widetilde{\chi}^0_1 \right)
\end{aligned}
\label{eqn:alt0a}
\end{equation}
can have sizeable cross sections for large values of $\tan\beta$ or $\mu > 0$. These two processes involve a cascade decay $\widetilde{\chi}^0_3 \rightarrow \widetilde{\chi}^0_2 \rightarrow \widetilde{\chi}^0_1$. This does not happen for smaller values of $\tan\beta$ and $\mu<0$ because the Higgsino splittings are smaller, keeping $m_{\widetilde{\chi}^0_2}$ nearly degenerate with $m_{\widetilde{\chi}^0_3}$. Finally, the process
\begin{equation}
pp \rightarrow \gamma ~ \left(\widetilde{\chi}^+ \rightarrow \widetilde{\chi}^0_1 j j^{\prime} \right) \left(\widetilde{\chi}^0_{2,3} \rightarrow \ell^+ \ell^- \widetilde{\chi}^0_1\right)
\label{eqn:alt3a}
\end{equation}
is another 2-3 production. Here the photon is again directly produced instead of being a decay product. The other two particles produced are a chargino which decays hadronicaly and a neutralino which goes through a leptonic decay.
			
\section{Minor cuts}
\label{sec:ExtraCuts}
In this section we motivate and explain some of the minor cut used in our analysis. Following the rest of the paper, the cuts are broken up into `small mass splitting' and `large mass splitting' scenarios.\\

{\bf Photon transverse momentum:} \\

\noindent `small mass' splitting (A and B): $30 \text{ GeV} < p_{T,\gamma} < 100 \text{ GeV}$ \\
`large mass' splitting (C and D): $45\text{ GeV} < p_{T, \gamma} < 135\text{ GeV}$ \\
	
The transverse momenta of the signal photon is determined by the mass splitting of the neutralinos as well as the boost of the parent particle. The lower bound removes the background from soft final-state radiation. After the leptonic angular cuts $\left|\Delta\phi_{\ell_1,\ell_2}\right|$ and $\left|\Delta\phi_{\ell\ell-\gamma}\right|$ have been established,  the surviving background events have the leptonic system recoiling off a hard initial-state radiation photon. The signal photon is not as hard as these, placing an upper bound.
\\

{\bf Missing energy magnitude and orientation:} \\

\noindent `small mass' splitting (A and B): \\
\,   $\slashed{E}_T < 65$ GeV, $ 0.2 < \left|\Delta_{\phi}(\slashed{E}_T, \gamma)\right| < 2.7$, and $ 1.0 < \left|\Delta_{\phi}(\ell\ell, \slashed{E}_T) \right| < 2.9$ \\
`large mass' splitting (C and D): no cut \\

In many searches for supersymmetry or Dark Matter, the expectation is for large amounts of missing transverse energy. However, in much of the well-forged parameter space, the dominant production is to nearly degenerate Higgsinos, $\widetilde{\chi}^0_{2,3}$, produced back-to-back. When they decay to the LSP, the net result is the two unobserved particles are approximately back-to-back.  The vector sum of the LSP momenta cancels to some degree, implying that the overall amount of $\slashed E_T$ will be small (at least compared to traditional supersymmetry searches). 

\begin{figure}[h!]\
\includegraphics[width= 3 in]{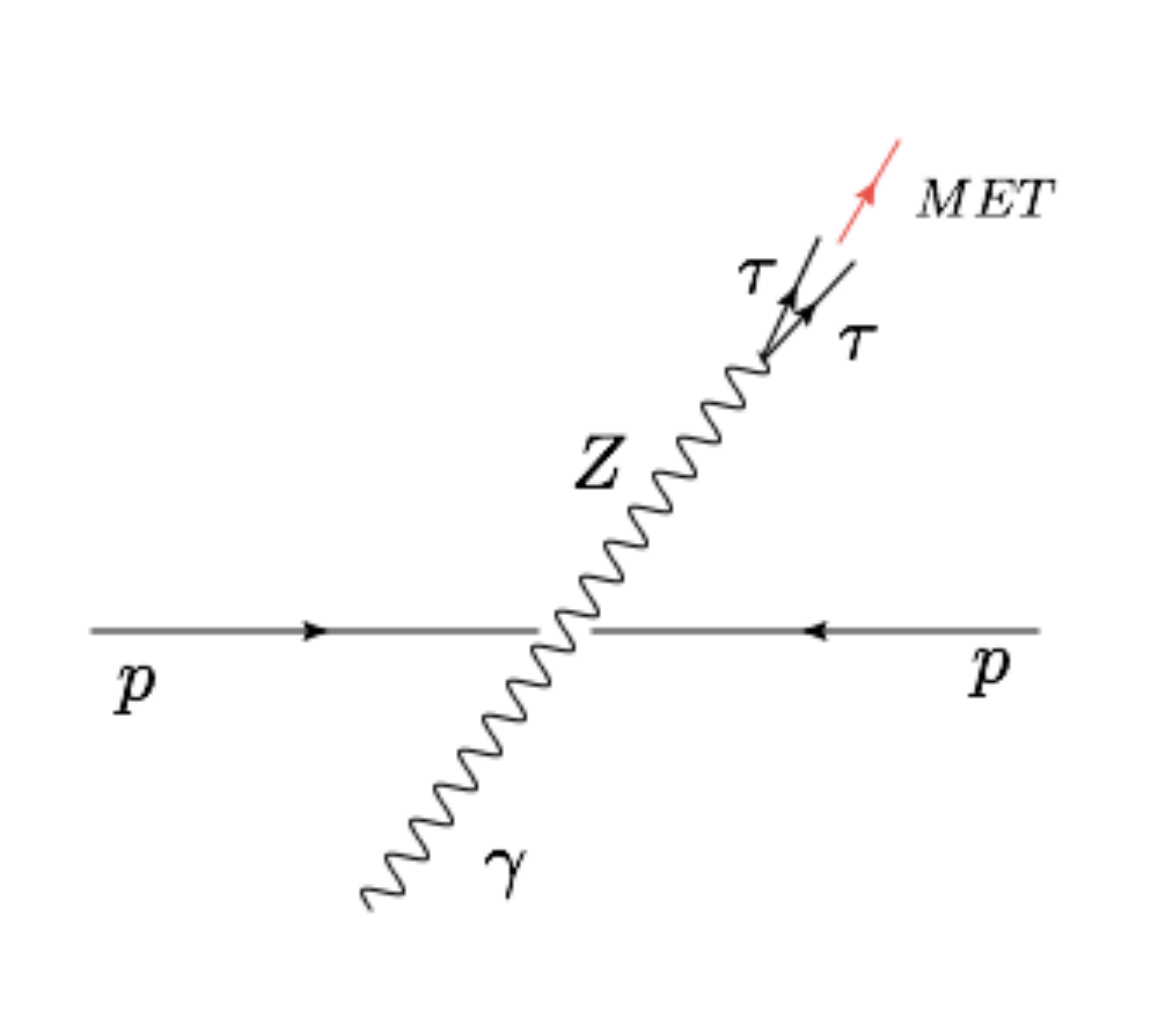}
\includegraphics[width= 3 in]{4}
\caption{The kinematics (after major cuts) of $\tau \tau \gamma \ (\widetilde{\chi}_2^0 \widetilde{\chi}_3^0)$ on left (right). In the  $\tau \tau \gamma$ picture, the azimuthal angle between the $\slashed E_T$ and the $\gamma$ is expected to be near $\pi$.  In the signal the $\slashed E_T$ vector will point neither towards the photon nor the dilepton system.}
\label{fig:TauBackground}
\end{figure}

The direction of the photon relative to the missing transverse energy can also help separate signal from background. In the $\gamma^*/Z(\tau^+\tau^-)+\gamma$ background, after demanding high $|\Delta \phi_{\ell\ell-\gamma}|$ and low $|\Delta \phi_{\ell\ell}|$, the surviving configurations have the neutrinos moving in the same direction as the leptons and in the opposite direction of the photon; thus, $|\Delta_{\phi}(\ell\ell, \slashed{E}_T)|$ is nearly $0$. The signal typically does not have this topology. Figure \ref{fig:TauBackground} shows example event configurations for the signal and background.

\bibliography{WellForged}
\bibliographystyle{JHEP}

\end{document}